\documentclass[aps,prb,reprint,groupedaddress,longbibliography]{revtex4-2}
\usepackage[T1]{fontenc}
\usepackage{amsmath,amssymb,mathtools,bm,mathrsfs}
\usepackage{graphicx,booktabs,array,etoolbox}
\usepackage{tikz}
\AtBeginEnvironment{thebibliography}{\raggedright}
\usepackage{microtype}
\usepackage[colorlinks=true,linkcolor=blue,citecolor=blue,urlcolor=blue]{hyperref}
\DeclareMathOperator{\Pf}{Pf}
\DeclareMathOperator{\Tr}{Tr}
\DeclareMathOperator{\tr}{tr}
\DeclareMathOperator{\Ind}{Ind}
\DeclareMathOperator{\SF}{SF}
\DeclareMathOperator{\USp}{USp}
\DeclareMathOperator{\Cl}{Cl}
\DeclareMathOperator{\Stab}{Stab}
\DeclareMathOperator{\Res}{Res}
\newcommand{\D}{\mathscr D}
\newcommand{\A}{\mathcal A}
\newcommand{\B}{\mathcal B}
\newcommand{\Z}{\mathbb Z}
\newcommand{\R}{\mathbb R}
\newcommand{\C}{\mathbb C}
\newcommand{\HH}{\mathbb H}
\newcommand{\one}{\mathbf 1}
\newcommand{\dd}{\mathrm d}
\newcommand{\slp}{\gamma^\mu\partial_\mu}

\newcommand{\cyl}{\widehat{\D}}
\newcommand{\cH}{\widehat H}

\begin{document}
\title{A Clifford-index approach connecting boundary Dirac fermions to sigma-model topological terms in the periodic table of topological insulators and superconductors}
\author{Y. X. Zhao}
\email{yuxinphy@hku.hk}
\affiliation{Department of Physics and HK Institute of Quantum Science \&
Technology, The University of Hong Kong, Pokfulam Road, Hong Kong, China}
\date{October 13, 2026}
\begin{abstract}
The periodic table of topological insulators and superconductors emerges from two distinct approaches: boundary Anderson localization and momentum-space K-theory. Connecting these approaches physically requires deriving the boundary sigma-model topological terms from the underlying fermion theories. We develop an index-theoretic framework for minimal massless boundary Dirac models in the eight real Altland--Zirnbauer classes. Its central ingredients are an elliptic cylinder operator interpolating between two sigma-field textures, a quotient index defined by the cylinder's graded Clifford symmetry algebra, and an index theorem identifying its protected zero-mode index with the relative homotopy class of the endpoint textures. To establish this correspondence, we deform the cylinder texture while preserving the kernel's index class and localize the obstruction to a Clifford defect. An exact oscillator-kernel calculation fixes the generator normalization. A field-independent congruence and a common regularization provide a real antisymmetric Pfaffian representative whose sign transport is related to the cylinder mod-two index by Witten's spectral-flow argument. This yields the nontrivial $\Z_2$ theta weight for one minimal physical module. Combining complex-parent anomalies with physical Clifford multiplicities and real-target periods also gives the real-class Wess--Zumino terms with unit spherical level. The framework thus provides an operator-level connection between boundary Dirac theories and the sigma-model topological terms associated with the periodic table.
\end{abstract}
\maketitle

\section{Introduction}
\label{sec:introduction}

Symmetry-protected topological phases have been a major focus of condensed matter physics since the discovery of time-reversal-invariant topological insulators~\cite{KaneMeleQSH,KaneMeleZ2}. A milestone was the periodic table of topological insulators and superconductors in the ten Altland--Zirnbauer (AZ) symmetry classes. These classes exhaust the symmetry-algebra possibilities generated by time-reversal and particle--hole symmetries within the tenfold framework~\cite{AltlandZirnbauer,Zirnbauer1996}. The periodic table was derived through two distinct approaches to characterizing topological stability: Anderson localization and momentum-space topological invariants~\cite{SRFL,RyuReview,Kitaev}.

Building on Anderson localization~\cite{Anderson1958} and its nonlinear sigma-model description~\cite{Wegner1979}, Schnyder, Ryu, Furusaki, and Ludwig considered massless boundary fermions subject to symmetry-preserving disorder. They argued that topologically nontrivial boundary fermions cannot be completely localized in the same way as topologically trivial fermions. This distinction can be reflected in topological terms of the nonlinear sigma models that describe the low-energy modes in the fermionic replica formulation. Each AZ class $a_{\mathrm{AZ}}$ corresponds to one of the ten classifying spaces of complex or real K-theory, denoted here by $R_{a_{\mathrm{AZ}}}$, in the stable range of the replica number. The possible topological terms are then determined by the homotopy groups $\pi_D(R_{a_{\mathrm{AZ}}})$, which are isomorphic to $0$, $\mathbb{Z}$, or $\mathbb{Z}_2$. An integer group $\pi_D(R_{a_{\mathrm{AZ}}})\cong\mathbb{Z}$ permits a Wess--Zumino (WZ) term in one lower dimension, $D-1$, whereas $\pi_D(R_{a_{\mathrm{AZ}}})\cong\mathbb{Z}_2$ permits a $\mathbb{Z}_2$ theta term in dimension $D$~\cite{SRFL,RyuReview}. The WZ level is quantized~\cite{WessZumino1971,WittenWZ1983}, and the $\mathbb{Z}_2$ theta weight is a discrete character. These coefficients cannot change continuously under renormalization-group flow, providing a topological basis for delocalization in the presence of symmetry-preserving disorder~\cite{SRFL,RyuAnomaly}. In this way, the classification of boundary topological terms across AZ classes and dimensions yields the periodic table of topological insulators and superconductors.

The table separates into a complex table, $\mathrm{Top}_{\mathbb{C}}(c,D)$, for the two complex AZ classes, with $c=1,2$, and a real table, $\mathrm{Top}_{\mathbb{R}}(s,D)$, for the eight real AZ classes, with $s=0,1,\ldots,7$. Both exhibit the so-called $(1,1)$ periodicity,
\begin{equation}
 \mathrm{Top}_{\mathbb{K}}(s,D)\cong
 \mathrm{Top}_{\mathbb{K}}(s+1,D+1),
\end{equation}
where $\mathbb{K}=\mathbb{C}$ or $\mathbb{R}$. In spatial dimension, the complex and real tables have periods two and eight, respectively:
\begin{equation}
 \begin{aligned}
 \mathrm{Top}_{\mathbb{C}}(c,D+2)&\cong\mathrm{Top}_{\mathbb{C}}(c,D),\\
 \mathrm{Top}_{\mathbb{R}}(s,D+8)&\cong\mathrm{Top}_{\mathbb{R}}(s,D).
 \end{aligned}
\end{equation}
These are the familiar periodicity patterns of complex and real K-theory~\cite{Bott1959,ABS,AtiyahReality}.

Kitaev gave an alternative derivation of the periodic table by classifying gapped momentum-space Hamiltonians $\mathcal H_a(\bm k)$ in the ten AZ classes using K-theory~\cite{Kitaev}. Kane and Mele had already connected the $\mathbb{Z}_2$ invariant to a mod-two index in twisted Real K-theory on the Brillouin torus, drawing on Atiyah and Singer's real index theory~\cite{KaneMeleZ2,AtiyahSingerReal}. Kitaev developed the K-theoretic classification for all AZ classes and dimensions within topological band theory. In this framework, gapped momentum-space Hamiltonians are classified by $\widetilde K^{-c}(S^D)$ in the complex classes and $\widetilde{KR}^{-s}(S^D)$ in the real classes, with an appropriate ordering of the symmetry classes. Here the $D$-dimensional momentum space is compactified to a sphere $S^D$; in the Real K-theory description, the momentum-reversal involution is part of the data~\cite{AtiyahReality,Kitaev}. This approach gives the same tables:
\begin{equation}
 \begin{aligned}
 \mathrm{Top}_{\mathbb{C}}(c,D)&\cong\widetilde K^{-c}(S^D),\\
 \mathrm{Top}_{\mathbb{R}}(s,D)&\cong\widetilde{KR}^{-s}(S^D).
 \end{aligned}
\end{equation}

The agreement between classifications derived from these two distinct physical considerations has a deeper origin, already apparent before the full periodic table was formulated. The surface states of a three-dimensional topological insulator are described by two-dimensional massless Dirac fermions~\cite{FuKaneMele3D,SRFL}. Whereas the bulk topology is characterized by a momentum-space K-theory invariant, the nonlinear sigma model for the surface Dirac fermions in the presence of time-reversal-invariant disorder contains a $\mathbb{Z}_2$ theta term~\cite{RyuAnomaly}. A complementary example for WZ terms is provided by Pruisken's nonlinear sigma model for the two-dimensional integer quantum Hall effect~\cite{Pruisken1984}. In a Hall plateau, the renormalized Hall conductivity approaches $\sigma_{xy}=k$ in units of $e^2/h$, corresponding to a topological coefficient $\theta=2\pi\sigma_{xy}\to2\pi k$, with $k\in\Z$. In a sample with a boundary, this quantized part of the bulk theta term survives as an edge action that can be written as a level-$k$ WZ term in the one-dimensional boundary sigma model describing the chiral edge fermions~\cite{PruiskenEdge1999,Pruisken2009}.

More generally, all topological insulators and superconductors in the two periodic tables admit massive Dirac representatives, whose in-gap boundary states are described by massless Dirac fermions~\cite{RyuReview,ZhaoWang2014,ZhaoWang2013}. A physical isomorphism between the two classifications can therefore be established by deriving the topological terms in the nonlinear sigma models for these disordered boundary fermions in every symmetry class and dimension. This task is comparatively straightforward in the complex classes: only WZ terms are involved, and they can be derived using perturbative variational methods based on parity and non-Abelian anomalies. The complex-class derivations have therefore already been carried out~\cite{ZhaoWang2015Response}; for class A, the WZ and related Chern--Simons formulations for disordered Weyl fermions are developed in Refs.~\cite{ZhaoWangWeyl2015,AltlandBagrets2015}. For the $\mathbb{Z}_2$ theta terms in the real classes, explicit low-dimensional derivations have used techniques tailored to particular models~\cite{Ostrovsky2007,RyuAnomaly}.

At the heart of our approach is a three-part index-theoretic construction. First, for a $D$-dimensional disordered massless boundary Dirac model in each real AZ class, we construct an elliptic $(D+1)$-dimensional cylinder operator by interpolating between two sigma-field textures. Second, we use the graded Clifford symmetry algebra of each cylinder operator to formulate a quotient index of its zero-mode space, modulo symmetry-compatible modules whose zero modes can be lifted. Third, we prove an index theorem identifying this zero-mode index with the relative homotopy class of the endpoint textures, including the generator normalization. This establishes an operator-level correspondence between the algebraic protection of cylinder zero modes and the topology of sigma-field textures. Combined with a regularized real-Pfaffian representation and Witten's spectral-flow argument~\cite{WittenSU2,Witten}, the theorem yields the nontrivial $\mathbb{Z}_2$ theta weights for the specified minimal models, providing the physical connection between the momentum-space and boundary-localization approaches to the classification.

\subsection{Structure of the manuscript}

The paper proceeds from the boundary fermion representation to the texture index and finally to the topological phase of the fermion integral. Section~\ref{sec:minimal-boundary} specifies the minimal boundary Dirac models and their physical Clifford modules. Section~\ref{sec:fermionic-action} introduces the disorder self-energy and fermionic kernels, and Sec.~\ref{sec:targets} derives the compact sigma-model targets.

Sections~\ref{sec:cylinder}--\ref{sec:quotients} construct the elliptic interpolation cylinder, derive its full graded symmetry algebra, and compute the module-extension index. The algebra depends only on $r=s-D-1\pmod8$, with quotient groups $\Z_2$ at $r=1,2$ and $\Z$ at $r=3,7$. Section~\ref{sec:texture-index-isomorphism} establishes the texture--index correspondence through index-preserving deformation and localization to a Clifford defect. Explicit $D=2$ examples illustrate the construction, and the oscillator kernel fixes the generator normalization.

Section~\ref{sec:congruence-pfaffian} constructs a field-independent congruence and the associated real antisymmetric representative, then specifies the common regularization of its Pfaffian. Section~\ref{sec:paired-spectral-flow} uses Witten's spectral-flow argument to derive the $\Z_2$ theta weight. Section~\ref{sec:wz-descent} combines the complex-parent anomaly with physical multiplicities and integer-index normalization to derive the real-class WZ terms one dimension lower. Section~\ref{sec:status} summarizes the physical connection and the scope of the results, including species counting and microscopic regulator matching.

\section{Massless boundary Dirac models}\label{sec:minimal-boundary}
\subsection{Physical setting and Clifford conventions}
We consider minimal noninteracting massless Dirac models in the eight real Altland--Zirnbauer classes. The physical bulk and boundary dimensions are related by
\begin{equation}
 d=D+1,\qquad D=\text{boundary spatial dimension}.
\end{equation}
This section fixes the physical kinetic matrices and symmetry representation before disorder and auxiliary sigma fields are introduced. Gamma matrices are Hermitian, and $K$ denotes complex conjugation in position space. We use $\Cl_{p,q}$ with $p$ anticommuting generators squaring to $+1$ and $q$ squaring to $-1$. Minimality refers to the full real physical Clifford module, not merely to the complex kinetic algebra. The construction does not assume that every such massless representative has a protected boundary charge~\cite{ZhaoWang2014,ZhaoWang2013,ABS,Kitaev}.

\subsection{Massless boundary Hamiltonian and physical symmetries}
\label{sec:boundary-dirac}
The starting point is a minimal massless boundary Dirac representative, not a bulk projection in a particular geometry. Following the systematic Dirac-matrix construction of boundary Fermi points in Refs.~\cite{ZhaoWang2014,ZhaoWang2013}, take one kinetic matrix per boundary spatial direction:
\begin{equation}\label{eq:minimal-boundary-H}
 \begin{aligned}
 H_0(\bm k)&=\sum_{\mu=1}^{D}k_\mu\gamma^\mu,\\
 (\gamma^\mu)^\dagger&=\gamma^\mu,\qquad
 \{\gamma^\mu,\gamma^\nu\}=2\delta^{\mu\nu}\one.
 \end{aligned}
\end{equation}
The size of these matrices is not chosen from the complex kinetic Clifford algebra alone: the physical antiunitary symmetries must act on the same spinor space. These works construct representatives and compute their charges; the real Clifford presentation below expresses the symmetry constraints in our conventions~\cite{ZhaoWang2014,ZhaoWang2013,ABS,Kitaev}.

Write physical time reversal and particle--hole symmetry as
\begin{equation}
 \mathcal T=u_TK,\qquad \mathcal C=u_CK,
 \qquad \mathcal T^2=\epsilon_T,\quad\mathcal C^2=\epsilon_C.
\end{equation}
The Bloch-Hamiltonian constraints, when the symmetry is present, are
\begin{equation}\label{eq:boundary-bloch-symmetry}
 \begin{aligned}
 u_T H_0(\bm k)^*u_T^{-1}
 &=H_0(-\bm k),\\
 u_C H_0(\bm k)^*u_C^{-1}
 &=-H_0(-\bm k).
 \end{aligned}
\end{equation}
Because $H_0$ is odd in momentum, comparison of each $k_\mu$ coefficient gives
\begin{equation}\label{eq:physical-gamma-constraints}
 \boxed{\begin{aligned}
 u_T\gamma^{\mu *}u_T^{-1}&=-\gamma^\mu,
 &u_Tu_T^*&=\epsilon_T,\\
 u_C\gamma^{\mu *}u_C^{-1}&=+\gamma^\mu,
 &u_Cu_C^*&=\epsilon_C.
 \end{aligned}}
\end{equation}
In position space complex conjugation also changes the factor $-i$ in $H_0=-i\gamma^\mu\partial_\mu$; hence the same relations give $u_TH_0^*u_T^{-1}=H_0$ and $u_CH_0^*u_C^{-1}=-H_0$, consistently with our later kernel conventions.

If both antiunitary symmetries are imposed, choose the phase of their product so that the physical chiral operator satisfies
\begin{equation}\label{eq:physical-Gamma}
 \begin{gathered}
 \Gamma^\dagger=\Gamma,\quad\Gamma^2=1,\quad
 \{\Gamma,\gamma^\mu\}=0,\qquad \mathcal C=\Gamma\mathcal T,\\
 \mathcal T\Gamma\mathcal T^{-1}=\kappa\Gamma,
 \qquad \kappa=\epsilon_T\epsilon_C.
 \end{gathered}
\end{equation}
The same sign gives $\mathcal C\Gamma\mathcal C^{-1}=\kappa\Gamma$. The eight real classes have
\begin{equation}\label{eq:physical-sign-table}
 \begin{array}{c|rrr}
 \text{class}&\epsilon_T&\epsilon_C&\kappa\\\hline
 \mathrm{AI}&+1&\text{--}&\text{--}\\
 \mathrm{AII}&-1&\text{--}&\text{--}\\
 \mathrm D&\text{--}&+1&\text{--}\\
 \mathrm C&\text{--}&-1&\text{--}\\
 \mathrm{BDI}&+1&+1&+1\\
 \mathrm{CII}&-1&-1&+1\\
 \mathrm{DIII}&-1&+1&-1\\
 \mathrm{CI}&+1&-1&-1
 \end{array}
\end{equation}
A dash means that the corresponding symmetry is not imposed. Accidental symmetries of a specially chosen clean Dirac matrix do not change the specified AZ constraints on the allowed perturbations.

\subsection{Real Clifford algebra of the physical Dirac model}
\label{sec:physical-clifford}
Regard the spinor space $V$ as a real vector space $V_\R$, and let $I$ denote multiplication by $i$. Then
\begin{equation}
\begin{gathered}
 I^2=-1,\qquad[I,\gamma^\mu]=[I,\Gamma]=0,\\
 I\mathcal T=-\mathcal T I,\qquad I\mathcal C=-\mathcal C I.
\end{gathered}
\end{equation}
On $V_\R$, antiunitary symmetries are real-linear operators. Equations~\eqref{eq:physical-gamma-constraints} become $\{\mathcal T,\gamma^\mu\}=0$ and $[\mathcal C,\gamma^\mu]=0$. Not all the original matrices mutually anticommute; the following combinations give genuine real Clifford generators. The construction of Clifford generators from symmetry operators and Dirac matrices is developed systematically by Morimoto and Furusaki~\cite{MorimotoFurusaki2013}. We retain the convention that $\Cl_{p,q}$ has $p$ generators of square $+1$ and $q$ of square $-1$, opposite to the ordering of the two signatures in that reference.

\paragraph{AI and AII.}
Use the pairwise anticommuting set
\begin{equation}\label{eq:physical-TR-generators}
 \{\gamma^1,\ldots,\gamma^D,\mathcal T,I\mathcal T\}.
\end{equation}
The kinetic generators have square $+1$, while $\mathcal T$ and $I\mathcal T$ both square to $\epsilon_T$. Thus AI gives $\Cl_{D+2,0}$ and AII gives $\Cl_{D,2}$.

\paragraph{D and C.}
Since $\mathcal C$ commutes with $\gamma^\mu$, use instead
\begin{equation}\label{eq:physical-PH-generators}
 \{I\gamma^1,\ldots,I\gamma^D,\mathcal C,I\mathcal C\}.
\end{equation}
These generators pairwise anticommute. Each $I\gamma^\mu$ has square $-1$, while $\mathcal C$ and $I\mathcal C$ have square $\epsilon_C$. The algebras are $\Cl_{2,D}$ for D and $\Cl_{0,D+2}$ for C.

\paragraph{BDI, CII, DIII, and CI.}
Start with the time-reversal generators in Eq.~\eqref{eq:physical-TR-generators} and adjoin
\begin{equation}\label{eq:physical-chiral-generator}
 F_\Gamma=
 \begin{cases}
 I\Gamma,&\kappa=+1,\quad F_\Gamma^2=-1,\\
 \Gamma,&\kappa=-1,\quad F_\Gamma^2=+1.
 \end{cases}
\end{equation}
In either case $F_\Gamma$ anticommutes with every kinetic generator and with both $\mathcal T$ and $I\mathcal T$. For example, when $\kappa=+1$, $\Gamma$ commutes with $\mathcal T$, but the extra $I$ supplies the required anticommutation. This gives $\Cl_{D+2,1}$ for BDI, $\Cl_{D,3}$ for CII, $\Cl_{D+1,2}$ for DIII, and $\Cl_{D+3,0}$ for CI.

Collecting the results, the \emph{physical} Dirac algebra $\mathfrak K_{\mathrm{AZ}}(D)$ and the complex dimension of its minimal physical spinor space are listed together below. Write $n_j=2^{\lfloor j/2\rfloor}$ and let $\beta_j$ be the eight-periodic factor equal to $1$ for $j\equiv0,1,2\pmod8$ and $2$ for $j\equiv3,4,5,6,7\pmod8$. Then
\begin{equation}\label{eq:physical-clifford-table}
 \boxed{\begin{array}{c|c|c}
 \text{class}&\mathfrak K_{\mathrm{AZ}}(D)&\dim_\C V_{\min}\\\hline
 \mathrm{AI}&\Cl_{D+2,0}&n_D\beta_{D+2}\\
 \mathrm{AII}&\Cl_{D,2}&n_D\beta_{D-2}\\
 \mathrm D&\Cl_{2,D}&n_D\beta_{2-D}\\
 \mathrm C&\Cl_{0,D+2}&n_D\beta_{-D-2}\\
 \mathrm{BDI}&\Cl_{D+2,1}&n_{D+1}\beta_{D+1}\\
 \mathrm{CII}&\Cl_{D,3}&n_{D+1}\beta_{D-3}\\
 \mathrm{DIII}&\Cl_{D+1,2}&n_{D+1}\beta_{D-1}\\
 \mathrm{CI}&\Cl_{D+3,0}&n_{D+1}\beta_{D+3}
 \end{array}}
\end{equation}
The last column counts the complex dimension of one irreducible physical module with its reconstructed complex structure, not the dimension of the algebra. Indeed, the irreducible real-module dimension of $\Cl_{p,q}$ is $2^{\lfloor(p+q)/2\rfloor}\beta_{p-q}$; dividing by two gives the displayed spinor dimensions. When the algebra has two simple summands, either irreducible module has this dimension: taking both is not part of the minimal-dimension count~\cite{ABS}.

These are presentations of the algebra generated by the imposed physical operators; they are not the cylinder commutant algebras $\mathfrak C^{(r)}$ computed later. An actual irreducible representation can have a nonfaithful image when the universal Clifford algebra has two simple summands.

The complex structure has not been added as an unconstrained extra generator. It is recovered from the two antiunitary-derived generators: if $a=\mathcal T$, $b=I\mathcal T$, then $I=\epsilon_Tba$, and similarly $I=\epsilon_Cba$ for $a=\mathcal C$, $b=I\mathcal C$. For the particle--hole presentation, the Hermitian kinetic matrices are recovered as $\gamma^\mu=-I(I\gamma^\mu)$. For the chiral presentation, recover $\Gamma=-IF_\Gamma$ when $\kappa=+1$, or $\Gamma=F_\Gamma$ when $\kappa=-1$, and then $\mathcal C=\Gamma\mathcal T$. Thus the Clifford module reconstructs the original symmetry-compatible complex Dirac theory, not just a list of square signs.

\subsection{Minimal representations, multiplicities, and model scope}
\label{sec:dirac-minimality}
A minimal Dirac representative is an irreducible real module $V_\R$ of the relevant algebra in Eq.~\eqref{eq:physical-clifford-table}, endowed with the reconstructed complex structure $I$. Choose the compatible positive inner product so that positive-square Clifford generators are orthogonal involutions and negative-square generators are orthogonal complex structures. The reconstructed $\gamma^\mu$ are then complex-linear Hermitian matrices, and the physical antiunitaries have the required squares. In particular,
\begin{equation}\label{eq:physical-spinor-size}
 \dim_\C V=\frac12\dim_\R V_\R.
\end{equation}
Multiple physical cones or species are direct sums of such modules. When there are two inequivalent irreducible modules, their sector labels must also be retained. For superconducting Hamiltonians this is the complex matrix size of the chosen BdG representation; the physical particle--hole redundancy must still be counted correctly in the Grassmann integral.

For bookkeeping across the eight classes and boundary dimensions, define
\begin{equation}\label{eq:rdefinition}
 \begin{gathered}
 \begin{array}{c|cccccccc}
 s&0&1&2&3&4&5&6&7\\\hline
 \text{class}&\mathrm{AI}&\mathrm{BDI}&\mathrm D&\mathrm{DIII}&\mathrm{AII}&\mathrm{CII}&\mathrm C&\mathrm{CI}
 \end{array}\\[3pt]
 r=s-D-1\pmod8.
 \end{gathered}
\end{equation}
Restrict the physical module to the complex algebra generated by the kinetic matrices and, in chiral classes, by $\Gamma$ as well. This gives $m=D$ or $D+1$ Hermitian Clifford generators. Dividing the minimal physical dimension in Eq.~\eqref{eq:physical-clifford-table} by the irreducible complex spinor dimension directly determines the total multiplicity:
\begin{equation}\label{eq:total-physical-multiplicity}
 \boxed{k=\frac{\dim_\C V_{\min}}{n_m},\qquad
 n_m=2^{\lfloor m/2\rfloor}.}
\end{equation}
Here $m=D$ in non-chiral classes and $m=D+1$ in chiral classes. For odd $m$, $k=k_++k_-$ includes both volume sectors. Table~\ref{tab:physical-multiplicities} distinguishes a quaternionic multiplicity from a pair of exchanged sectors. Along every fixed-$r$ diagonal, both the multiplicity and its sector interpretation agree. The physical intertwiner algebra and antiunitary action are derived in Sec.~\ref{sec:boundary-intertwiners}; Sec.~\ref{sec:full-symmetry} later supplies their cylinder lift. The pattern is eight-periodic in $D$: both numerator and denominator increase by 16 under $D\mapsto D+8$.

\begin{table*}[t]\centering\small
\caption{Total multiplicity $k=\dim_\C V_{\min}/2^{\lfloor m/2\rfloor}$ from Eq.~\eqref{eq:physical-clifford-table}, with $m=D$ (non-chiral) or $D+1$ (chiral). An unmarked $2$ is a quaternionic multiplicity within one complex Clifford sector. The marked $2_{\mathrm{ex}}$ instead means $k_+=k_-=1$ in opposite volume sectors exchanged by the physical antiunitary, not multiplicity two in each sector.}
\label{tab:physical-multiplicities}
\begin{tabular}{@{}lcccccccc@{}}
\toprule
Class&$D=0$&$D=1$&$D=2$&$D=3$&$D=4$&$D=5$&$D=6$&$D=7$\\\midrule
AI&$1$&$2_{\mathrm{ex}}$&$2$&$2$&$2$&$2_{\mathrm{ex}}$&$1$&$1$\\
BDI&$1$&$1$&$2_{\mathrm{ex}}$&$2$&$2$&$2$&$2_{\mathrm{ex}}$&$1$\\
D&$1$&$1$&$1$&$2_{\mathrm{ex}}$&$2$&$2$&$2$&$2_{\mathrm{ex}}$\\
DIII&$2_{\mathrm{ex}}$&$1$&$1$&$1$&$2_{\mathrm{ex}}$&$2$&$2$&$2$\\
AII&$2$&$2_{\mathrm{ex}}$&$1$&$1$&$1$&$2_{\mathrm{ex}}$&$2$&$2$\\
CII&$2$&$2$&$2_{\mathrm{ex}}$&$1$&$1$&$1$&$2_{\mathrm{ex}}$&$2$\\
C&$2$&$2$&$2$&$2_{\mathrm{ex}}$&$1$&$1$&$1$&$2_{\mathrm{ex}}$\\
CI&$2_{\mathrm{ex}}$&$2$&$2$&$2$&$2_{\mathrm{ex}}$&$1$&$1$&$1$\\
\bottomrule
\end{tabular}
\end{table*}

For $D=2$ these algebras predict
\begin{equation}\label{eq:d2-minimal-sizes}
 \begin{array}{c|c|c}
 \text{class}&\mathfrak K_{\mathrm{AZ}}(2)&\dim_\C V_{\min}\\\hline
 \mathrm{AI}&\Cl_{4,0}\simeq M_2(\HH)&4\\
 \mathrm{AII}&\Cl_{2,2}\simeq M_4(\R)&2\\
 \mathrm D&\Cl_{2,2}\simeq M_4(\R)&2\\
 \mathrm C&\Cl_{0,4}\simeq M_2(\HH)&4\\
 \mathrm{BDI}&\Cl_{4,1}\simeq M_4(\C)&4\\
 \mathrm{CII}&\Cl_{2,3}\simeq M_4(\C)&4\\
 \mathrm{DIII}&\Cl_{3,2}\simeq M_4(\R)^{\oplus2}&2\\
 \mathrm{CI}&\Cl_{5,0}\simeq M_2(\HH)^{\oplus2}&4
 \end{array}
\end{equation}
The later worked examples in Sec.~\ref{sec:d2-examples} realize exactly these dimensions. The original physical representation determines the multiplicity and sector-exchange intertwiners that enter the cylinder symmetry calculation; the auxiliary field $Q$ and its pairing matrix $b$ have not entered $\mathfrak K_{\mathrm{AZ}}(D)$.

Irreducibility establishes the smallest model obeying the imposed symmetries; it does not by itself prove a nontrivial boundary topological charge. A normalized Hermitian physical mass would obey
\begin{equation}
 M^2=1,\quad\{M,\gamma^\mu\}=0,\qquad
 \begin{aligned}
 u_TM^*u_T^{-1}&=+M,\\
 u_CM^*u_C^{-1}&=-M,
 \end{aligned}
\end{equation}
with $\{M,\Gamma\}=0$ when chiral symmetry is imposed. Its existence is a further symmetry-extension question. Accordingly, a minimal representative can be topologically trivial, and extra copies can cancel a protected charge. Refs.~\cite{ZhaoWang2014,ZhaoWang2013} supply the Fermi-point charge and bulk--boundary matching for topological representatives, including the integer and two descendant $\Z_2$ types; they do not assert that every irreducible kinetic model is protected. The sphere used to define a boundary Fermi-point charge is in momentum/frequency space; it is distinct from the real-space texture compactification introduced in Sec.~\ref{sec:fermionic-action}.

After the physical intertwiner analysis below, Sec.~\ref{sec:fermionic-action} introduces symmetry-preserving disorder and its auxiliary self-energy. That self-energy is not the Hermitian physical mass just described. The distinction keeps the boundary Dirac model separate from the auxiliary kernel and its later cylinder symmetry algebra.

\subsection{Symmetry intertwiners of the minimal physical modules}
\label{sec:boundary-intertwiners}
Before introducing auxiliary fields, we determine the algebra of physical intertwiners and the action of the imposed antiunitary. This is distinct from the Clifford algebra $\mathfrak K_{\mathrm{AZ}}(D)$ generated by the kinetic and symmetry operators themselves. For the generators
\begin{equation}\label{eq:physical-E}
 \begin{gathered}
 (E_1,\ldots,E_m)=
 \begin{cases}
 (\gamma^1,\ldots,\gamma^D),&\text{non-chiral},\\
 (\gamma^1,\ldots,\gamma^D,\Gamma),&\text{chiral},
 \end{cases}\\
 m=D\ \text{(non-chiral)},\qquad m=D+1\ \text{(chiral)},
 \end{gathered}
\end{equation}
set
\begin{equation}\label{eq:graded-commutant}
 \mathcal E_p=\{a\in\operatorname{End}_{\C}V:
 aE_j=(-1)^pE_ja\ \forall j\},\qquad p=0,1.
\end{equation}
The complex-linear intertwiner algebra is $\mathcal E_0\oplus\mathcal E_1$. Its even and odd elements commute and anticommute, respectively, with $H_0$; in chiral classes we also require the same sign for $\Gamma$. Thus we retain precisely the physical intertwiners compatible with the later chiral self-energy, rather than accidental symmetries of the kinetic operator alone. Write $\Theta=uK$, choosing physical time reversal when present and particle--hole symmetry in D and C. Its square and its action on every $E_j$ are fixed by Eqs.~\eqref{eq:physical-gamma-constraints} and \eqref{eq:physical-Gamma}. Together with $\Theta$, the intertwiner spaces generate a real-linear operator algebra. The following lists specify it concretely for every minimal model.

\paragraph{Even number of physical generators.}
For even $m$ ($r$ odd), the complex Clifford spinor occurs with multiplicity $k=1$ at $r=1,7$ and $k=2$ at $r=3,5$. Hence
\begin{equation}\label{eq:even-commutant}
 \begin{gathered}
 \mathcal E_0=\one\otimes M_k(\C),\qquad
 \mathcal E_1=\omega_m\mathcal E_0,\\
 \omega_m=(-i)^{m/2}E_1\cdots E_m,\qquad \omega_m^2=1.
 \end{gathered}
\end{equation}
Set $\omega_0=1$. When the single-sector multiplicity is two, $\Theta$ acts on that factor through $i\rho_yK$. The physical multiplicity generators are
\begin{equation}\label{eq:physical-multiplicity-seeds}
 \begin{gathered}
 j_a=\one\otimes i\rho_a,\qquad a=x,y,z,\\
 j_a^2=-1,\qquad j_xj_y=-j_z,\qquad
 [j_a,E_j]=[j_a,\Theta]=0.
 \end{gathered}
\end{equation}
Distinct $j_a$ anticommute, and these generators commute with $i\one$ and with $\omega_m$ when it exists. They are required by the physical representation, not optional extra species.

\paragraph{Odd number of physical generators.}
For odd $m$, a single volume sector occurs at $r=0$ with $k=1$ or at $r=4$ with $k=2$. In these cases $\mathcal E_0=\one\otimes M_k(\C)$ and $\mathcal E_1=0$; the $j_a$ occur only at $r=4$. At $r=2,6$, one copy of each opposite volume sector is required instead. Choose
\begin{equation}\label{eq:physical-sector-exchange}
 E_j=\rho_z\otimes e_j,\qquad
 \ell=\rho_z\otimes1,\qquad a_{\mathrm{ex}}=\rho_x\otimes1.
\end{equation}
Then
\begin{equation}\label{eq:odd-commutant}
 \begin{aligned}
 \mathcal E_0&=\operatorname{span}_\C\{1,\rho_z\}\otimes1,\\
 \mathcal E_1&=\operatorname{span}_\C\{\rho_x,\rho_y\}\otimes1.
 \end{aligned}
\end{equation}
Here $\ell^2=a_{\mathrm{ex}}^2=1$, $\ell a_{\mathrm{ex}}=-a_{\mathrm{ex}}\ell$, and $\Theta\ell\Theta^{-1}=-\ell$. Thus $i\one,\Theta$, together with the available $j_a$, odd intertwiner $\omega_m$, or sector operators $\ell,a_{\mathrm{ex}}$, determine the physical operator algebra in all eight cases.

\paragraph{Antiunitary action on the odd intertwiners.}
Write $\Theta^2=\epsilon$ and $\Theta E_j\Theta^{-1}=\eta_jE_j$. Set $h=\lfloor m/2\rfloor$ and let $\nu$ count the negative $\eta_j$. Conjugating the normalized volume gives
\begin{equation}\label{eq:r-only-volume-sign}
 \begin{gathered}
 \Omega_m=(-i)^hE_1\cdots E_m,\\
 \Theta\Omega_m\Theta^{-1}=\chi\Omega_m,
 \qquad \chi=(-1)^{h+\nu}.
 \end{gathered}
\end{equation}
For even $m$, this is the action on $\omega_m$. For odd $m$, the physical signs give $\chi=+1$ at $r=0,4$ and $-1$ at $r=2,6$, explaining preservation or exchange of the volume sectors. In the exchanged cases choose a single-sector intertwiner $u_0e_j^*u_0^{-1}=-\eta_je_j$. Its square is fixed by the real or quaternionic type of $\Cl_{\nu,m-\nu}$, whose signature difference is $2\nu-m\equiv1$ or $5\pmod8$. Therefore
\begin{equation}\label{eq:r-only-exchange-sign}
 \begin{gathered}
 \epsilon_0=u_0u_0^*=(-1)^{(\nu-h-1)/2},\\
 \Theta a_{\mathrm{ex}}\Theta^{-1}=\epsilon\epsilon_0a_{\mathrm{ex}}.
 \end{gathered}
\end{equation}
The exponent is integral because $\chi=-1$. The sector factor of $u$ is $\rho_x$ if $\epsilon=\epsilon_0$ and $i\rho_y$ otherwise, which verifies the second relation. These are physical-module statements; no auxiliary pairing or cylinder Pauli matrix has yet been introduced. Section~\ref{sec:full-symmetry} promotes them to symmetries of the entire cylinder family.

\subsection{Two-dimensional physical realizations}
\label{sec:d2-examples}
For $D=2$, Eq.~\eqref{eq:physical-clifford-table} fixes all physical spinor sizes. In the non-chiral and single-sector chiral cases choose
\begin{equation}\label{eq:d2-gammas}
 \gamma^1=\sigma_x\otimes1_k,\qquad
 \gamma^2=\sigma_y\otimes1_k.
\end{equation}
AII and D have $k=1$, while AI and C have $k=2$. The physical antiunitary matrices are respectively
\begin{equation}
 \begin{aligned}
 u_{\mathrm{AII}}&=i\sigma_y,&u_{\mathrm D}&=\sigma_x,\\
 u_{\mathrm{AI}}&=i\sigma_y\otimes i\rho_y,&
 u_{\mathrm C}&=\sigma_x\otimes i\rho_y.
 \end{aligned}
\end{equation}
Here the physical odd seed is $\omega_2=\sigma_z\otimes1_k$, and $j_a=1_\sigma\otimes i\rho_a$ when $k=2$. These give $r=1,7,5,3$, respectively. DIII and CI use the same kinetic matrices with $\Gamma=\sigma_z\otimes1_k$, $k=1,2$, and the AII/AI time-reversal matrices; their single chiral volume sector gives $r=0,4$ with no odd generator.

For CII and BDI use the paired-sector convention
\begin{equation}\label{eq:d2-paired-gammas}
 \begin{gathered}
 \gamma^{1,2}=\rho_z\otimes\sigma_{x,y},\qquad
 \Gamma=\rho_z\otimes\sigma_z,\\
 u_{\mathrm{CII}}=i\rho_y\otimes\sigma_x,\qquad
 u_{\mathrm{BDI}}=\rho_x\otimes\sigma_x,\\
 \ell=\rho_z\otimes1,\qquad a_{\mathrm{ex}}=\rho_x\otimes1.
 \end{gathered}
\end{equation}
These paired realizations have $r=2,6$, respectively. Together with the preceding cases, they explicitly realize all eight physical intertwiner algebras at $D=2$; their cylinder promotion is performed in Sec.~\ref{sec:full-symmetry}.

\section{Disordered Dirac fermions and the sigma-model kernel}\label{sec:fermionic-action}
\subsection{Disorder averaging, auxiliary-space conventions and the effective sigma model action}
Add symmetry-preserving disorder to the boundary Hamiltonian $H_0$ of Eq.~\eqref{eq:minimal-boundary-H}, 
\begin{equation}\label{eq:boundary-position-H}
	H_0(-i\nabla)=-i\gamma^\mu\partial_\mu.
\end{equation}
Replica averaging and Hubbard--Stratonovich decoupling introduce an auxiliary matrix field $Q$; its fixed-amplitude saddle manifold becomes the compact nonlinear sigma-model target~\cite{Wegner1979,SRFL,Koenig}. 


Because of the antiunitary symmetry in each AZ class, the auxiliary doubling is chosen so that the independent real Grassmann variables $\chi$ give
\begin{equation}\label{eq:action}
	S[\chi,Q]=S_{\mathrm{HS}}[Q]
	+\frac12\int\dd^Dx~
	\chi^T(x)\A[Q]\chi(x).
\end{equation}
Before freezing radial modes, a Gaussian disorder ensemble gives a quadratic Hubbard--Stratonovich penalty (with ensemble-dependent coefficients). On the fixed-amplitude saddle it is constant or supplies source terms. No universal numerical coefficient of $S_{\mathrm{HS}}$ is assumed here. The existence of a diffusive saddle and a controlled gradient expansion remains an assumption of the sigma-model description~\cite{SRFL,RyuReview,Koenig}. The saddle constraints are imposed only after the auxiliary-field formulation has been specified.

The disorder self-energy defines the Dirac-$Q$ kernel
\begin{equation}
	\boxed{\A[Q]=\B^{-1}\D[Q]}
\end{equation}
with
\begin{equation}\label{eq:kernel-general}
	\D[Q]=\slp+\Delta M(Q),\quad \B=u\otimes b.
\end{equation}
The physical matrices $u,\gamma^\mu,\Gamma$ and auxiliary matrices $b,Q$ act on separate tensor factors; identity factors are suppressed when unambiguous. For non-chiral classes $u=u_T$ or $u_C$, whereas for chiral classes $u=u_T$. $M(Q)$ will be specified later for non-chiral and chiral classes, respectively. Here $\Delta>0$ sets its amplitude and $\B$ supplies the auxiliary pairing.

Although $H_0$ is Hermitian, $\D$ is generally not. Grassmann antisymmetry requires $\A^{\mathsf t}=-\A$, not necessarily $\A^\dagger=-\A$ or $\A^*=\A$. We adopt the convention that the superscript $T$ transposes internal matrices, whereas $\mathsf t$ denotes the full operator transpose, including integration by parts:
\begin{equation}
 (\partial_\mu)^{\mathsf t}=-\partial_\mu,
 \qquad (\partial_\mu)^\dagger=-\partial_\mu.
\end{equation}
All adjoint and symmetry identities require compatible operator domains.

For the later texture and index calculations, impose $Q(x)\to Q_*$ at spatial infinity, giving a based map $S^D\to\mathcal M$. This compactification is in real space. Constant gamma matrices describe flat space; compatible curved or infrared-regularized realizations require covariant derivatives, parallel Clifford structures, and the corresponding domain conditions.

With a common ultraviolet regulator and consistent orientation of the fermionic measure, the integration over $\chi$ leads to 
\begin{align}
	e^{-S_{\mathrm{eff}}[Q]}&=e^{-S_{\mathrm{HS}}[Q]}F[Q] ,
\end{align}
with 
\begin{equation}
	F[Q]=\Pf_{\mathrm{reg}}\A[Q].
\end{equation}
Restricting $Q$ to its normalized orbit and integrating out massive radial fluctuations yields the nonlinear sigma model. A long-wavelength expansion of this effective action supplies its gradient terms, with coefficients set by the disorder ensemble and regulator. 


In the non-chiral classes, 
\begin{equation}\label{eq:nonchiral-embedding}
 M(Q)=Q,\qquad Q^\dagger=Q,\qquad Q^2=1,
\end{equation}
i.e., $Q$ is a Hermitian involution.
Therefore,
\begin{equation}
	\D[Q]=\slp+\Delta Q.
\end{equation}

For the chiral classes use the physical $\Gamma$ from Eq.~\eqref{eq:physical-Gamma}. The sigma field is unitary, not a Hermitian involution:
\begin{equation}\label{eq:chiralD}
 Q^\dagger Q=1,\quad
 M(Q)=P_+Q+P_-Q^\dagger,\quad P_\pm=\frac{1\pm\Gamma}{2},
\end{equation}
and therefore
\begin{equation}
 \boxed{\D[Q]=
 \begin{pmatrix}\Delta Q&\partial_-\\\partial_+&\Delta Q^\dagger\end{pmatrix}}.
\end{equation}
Here $\partial_-=P_+\slp P_-$ and $\partial_+=P_-\slp P_+$ denote the restricted off-diagonal kinetic blocks. The two diagonal blocks are the conjugate chiral self-energies. This structure is explicit in the chiral disorder construction~\cite{Koenig}.





\section{Sigma-model target manifolds}\label{sec:targets}
The kernels $\mathcal{A}[Q]$ determine the target through two distinct steps: Grassmann antisymmetry fixes the allowed auxiliary pairing and linear constraints, and the normalized saddle fixes the orbit of $Q$. We carry this out for the non-chiral and chiral real classes in turn.

\subsection{Non-chiral classes: AI, AII, D, C}\label{sec:nonchiral}
\paragraph{Physical input and auxiliary pairing.}
Let the defining antiunitary be $uK$, with
\begin{equation}\label{eq:nonphysical}
 uH_0^*u^{-1}=\varsigma H_0,\qquad uu^*=\epsilon\one.
\end{equation}
Thus $\varsigma=+1$ for time reversal and $\varsigma=-1$ for particle--hole symmetry, and
\begin{equation}
 u^T=\epsilon u,\qquad
 u\gamma^{\mu T}u^{-1}=-\varsigma\gamma^\mu.
\end{equation}
The non-chiral embedding is Eq.~\eqref{eq:nonchiral-embedding}; its condition $Q^2=1$ is a saddle constraint, not an AZ symmetry identity.

\paragraph{Linear constraints from fermionic antisymmetry.}
The kinetic term of $\A$ is antisymmetric precisely with the auxiliary transpose parity
\begin{equation}\label{eq:bnon}
 \boxed{b^T=-\varsigma\epsilon b}.
\end{equation}
For example, $(u^{-1}\slp)^{\mathsf t}=\varsigma\epsilon u^{-1}\slp$; the auxiliary bilinear must contribute the opposite sign. Applying the same requirement to the self-energy yields
\begin{equation}\label{eq:qnon}
 \boxed{bQ^Tb^{-1}=\varsigma Q}.
\end{equation}
Consequently,
\begin{equation}\label{eq:Dnon}
 \B^T=-\varsigma\B,\qquad
 \B\D^{\mathsf t}\B^{-1}=\varsigma\D,
 \qquad (\B^{-1}\D)^{\mathsf t}=-\B^{-1}\D.
\end{equation}
These are transpose constraints on a non-Hermitian kernel, not the original Hamiltonian's antiunitary equation at fixed $Q$.

\begin{table}[tb]\centering\small
\caption{Non-chiral input data. Every row also has $Q^\dagger=Q$, $Q^2=1$.}
\label{tab:nonchiral-input}
\begin{tabular}{@{}lcccl@{}}
\toprule
Class&$\varsigma$&$\epsilon$&$b$&Constraint on $Q$\\\midrule
AI &$+1$&$+1$&$J$&$JQ^TJ^{-1}=Q$\\
AII&$+1$&$-1$&$1$&$Q^T=Q$\\
D  &$-1$&$+1$&$1$&$Q^T=-Q$\\
C  &$-1$&$-1$&$J$&$JQ^TJ^{-1}=-Q$\\\bottomrule
\end{tabular}
\end{table}

\paragraph{The normalized saddle orbit.}
Define
\begin{equation}
\begin{gathered}
 G_b=\{g\in U(L):gbg^T=b\},\\
 Q=gQ_*g^{-1},\qquad
 \mathcal M=G_b/\Stab_{G_b}(Q_*).
\end{gathered}
\end{equation}
For symmetric $b$, $G_b$ is orthogonal; for antisymmetric $b$, it is compact symplectic. The stabilizers follow from the eigenspaces of $Q_*$:
\begin{itemize}
\item AI: the positive and negative eigenspaces are quaternionic. Choose $b=\operatorname{diag}(J_N,J_N)$ and $Q_*=\operatorname{diag}(1_{2N},-1_{2N})$.
\item AII: $Q$ is real symmetric; the two eigenspaces are real $N$-planes.
\item D: $Q=iA$ with $A$ real, $A^T=-A$, $A^2=-1$. Such $A$ is an orthogonal complex structure, stabilized by $U(N)$.
\item C: $JK$ exchanges the two eigenspaces. The stabilizer consists of matrices $\operatorname{diag}(V,V^*)$ with $V\in U(N)$.
\end{itemize}
Here, the auxiliary antisymmetric form is given by
\begin{equation}
	J_N=\begin{bmatrix}0&\one_N\\-\one_N&0\end{bmatrix},
	\qquad J_N^T=-J_N,\qquad J_N^2=-\one.
\end{equation}
Therefore~\cite{Zirnbauer1996,SRFL,RyuReview},
\begin{equation}\label{eq:non-targets}
\begin{aligned}
\mathcal M_{\mathrm{AI}}&=\USp(4N)/[\USp(2N)\times\USp(2N)],\\
\mathcal M_{\mathrm{AII}}&=O(2N)/[O(N)\times O(N)],\\
\mathcal M_{\mathrm D}&=O(2N)/U(N),\\
\mathcal M_{\mathrm C}&=\USp(2N)/U(N).
\end{aligned}
\end{equation}
We write $\USp(2N)$ for the complex $2N\times2N$ unitary symplectic group. 

\subsection{Chiral classes: BDI, CII, DIII, CI}\label{sec:chiral}
\paragraph{Physical input and chiral embedding.}
Introduce $\Gamma^\dagger=\Gamma$, $\Gamma^2=1$, $\{\Gamma,\gamma^\mu\}=0$. Choose $u$ as the unitary part of physical time reversal:
\begin{equation}
 uu^*=\epsilon_T,\qquad
 u\gamma^{\mu *}u^{-1}=-\gamma^\mu,\qquad
 u\Gamma^*u^{-1}=\kappa\Gamma.
\end{equation}
With $\mathcal C=(\Gamma u)K$,
\begin{equation}
 \epsilon_C=\mathcal C^2=\kappa\epsilon_T,
 \qquad \kappa=\epsilon_T\epsilon_C.
\end{equation}
The chiral kernel and its normalized self-energy are given in Eq.~\eqref{eq:chiralD}.

\paragraph{Constraints and target.}
Fermionic antisymmetry requires
\begin{equation}
 b^T=-\epsilon_Tb,\qquad
 bQ^Tb^{-1}=
 \begin{cases}Q,&\kappa=+1,\\Q^\dagger,&\kappa=-1.\end{cases}
\end{equation}
Indeed, $uP_\pm^Tu^{-1}=P_{\pm\kappa}$: time reversal either preserves or exchanges the two chiral sectors. In every row,
\begin{equation}\label{eq:Dchiral-sym}
 \boxed{\B^T=-\B,\qquad \B\D^{\mathsf t}\B^{-1}=\D,
 \qquad \A^{\mathsf t}=-\A}.
\end{equation}
\begin{table}[tb]\centering\small
\caption{Chiral input data; all endpoint fields obey $Q^\dagger Q=1$.}
\label{tab:chiral-input}
\begin{tabular}{@{}lccccl@{}}
\toprule
Class&$\epsilon_T$&$\epsilon_C$&$\kappa$&$b$&Constraint\\\midrule
BDI &$+1$&$+1$&$+1$&$J$&$JQ^TJ^{-1}=Q$\\
CII &$-1$&$-1$&$+1$&$1$&$Q^T=Q$\\
DIII&$-1$&$+1$&$-1$&$1$&$Q^T=Q^\dagger$\\
CI  &$+1$&$-1$&$-1$&$J$&$JQ^TJ^{-1}=Q^\dagger$\\\bottomrule
\end{tabular}
\end{table}

For BDI, $QJ$ is antisymmetric unitary and $Q=VJV^TJ^{-1}$. For CII, $Q$ is symmetric unitary and $Q=VV^T$. The redundancies in $V$ are respectively $\USp(2N)$ and $O(2N)$. For DIII, the constraint is $Q^*=Q$, while for CI it is $QJQ^T=J$. Thus~\cite{RyuReview,Koenig},
\begin{equation}\label{eq:chiral-targets}
\begin{aligned}
\mathcal M_{\mathrm{BDI}}&=U(2N)/\USp(2N),\\
\mathcal M_{\mathrm{CII}}&=U(2N)/O(2N),\\
\mathcal M_{\mathrm{DIII}}&=O(2N),\\
\mathcal M_{\mathrm{CI}}&=\USp(2N).
\end{aligned}
\end{equation}
The group targets can also be written $(G_L\times G_R)/G_{\mathrm{diag}}$.

The physical chiral and particle--hole relations act covariantly on the self-energy family, not as anticommutation at fixed $Q$:
\begin{equation}
\begin{gathered}
 \Gamma\D[Q]\Gamma=-\D[-Q],\\
 \B_C\D[Q]^{\mathsf t}\B_C^{-1}=-\D[-Q],\quad  \B_C=(\Gamma u)\otimes b.
\end{gathered}
\end{equation}
In particular, $\B_C^T=-\kappa\B_C$.

It is noteworthy that in this work homotopy groups are evaluated at sufficiently large integer auxiliary rank in the so-called stable regimes.

\section{The elliptic interpolation cylinder}\label{sec:cylinder}
The central construction is an elliptic operator associated with an interpolation of the Dirac--$Q$ kernel. Its zero modes will carry the symmetry-module index that compares endpoint textures.
\subsection{Interpolation and asymptotic conditions}
Choose endpoint fields $Q_0,Q_1$ and interpolate in the larger Hubbard--Stratonovich space:
\begin{equation}\label{eq:linearpath}
 Q_t=(1-t)Q_0+tQ_1.
\end{equation}
Different homotopy sectors cannot be joined within $\mathcal M$, but this path preserves the real-linear symmetry constraints. It need not obey $Q_t^2=1$ or $Q_t^\dagger Q_t=1$. Both mass embeddings are real-linear, so $\A_t=(1-t)\A_0+t\A_1$ remains antisymmetric. The interpolation therefore remains in the allowed fermionic-kernel family even when it leaves the normalized target.

The auxiliary coordinate $t$ is not physical time; the cylinder dimension is $D+1$. Extend the path constantly outside $[0,1]$ and use the infinite cylinder with invertible asymptotic operators, or specify symmetry-compatible boundary conditions on a finite cylinder. No identification of the two endpoint fields is assumed.

\subsection{Operator, elliptic symbol, and mass coefficients}
Define
\begin{equation}\label{eq:cylinder}
 \boxed{\cyl=
 \begin{bmatrix}\partial_t&\D_t\\-\D_t^\dagger&-\partial_t\end{bmatrix},
 \qquad \cH=i\cyl.}
\end{equation}
Then $\cyl^\dagger=-\cyl$. Its principal symbol is
\begin{equation}
\begin{aligned}
 \sigma(\cyl)(\tau,\xi)&=i\bigl(\tau_3\tau+\tau_1\gamma^\mu\xi_\mu\bigr),\\
 \sigma(\cyl)^2&=-(\tau^2+|\xi|^2)\one.
\end{aligned}
\end{equation}
Thus the extra Pauli/Clifford factor makes the cylinder elliptic. The naive operator $\partial_t+\D_t$ lacks this property in our Euclidean conventions.

For non-chiral classes,
\begin{equation}\label{eq:cyl-non}
 \cyl=\tau_3\partial_t+\tau_1\slp+i\tau_2\Delta Q_t.
\end{equation}
For chiral classes, put $R_t=(Q_t+Q_t^\dagger)/2$, $S_t=(Q_t-Q_t^\dagger)/(2i)$. Then
\begin{equation}\label{eq:cyl-ch}
 \boxed{\cyl=\tau_3\partial_t+\tau_1\slp
 +i\tau_2\Delta R_t+i\tau_1\Gamma\Delta S_t.}
\end{equation}
The last term must be included when checking the grading symmetry.

\section{Lifting the boundary symmetry algebras to the cylinder}
\label{sec:full-symmetry}
The purpose of this section is to promote the physical symmetry intertwiners derived in Sec.~\ref{sec:boundary-intertwiners} to the complete graded internal symmetry algebra of the cylinder operator $\cH$. The central conclusion is that this algebra depends on the AZ class and boundary dimension only through $r$:
\begin{equation}\label{eq:r-only-algebra}
 \boxed{\mathfrak C_{\mathrm{AZ}}(D)\cong_{\mathrm{gr}}
 \mathfrak C^{(r)},\qquad r=s-D-1\pmod8.}
\end{equation}
For a minimal physical model, the complex-linear graded intertwiner algebra $\mathcal E_0\oplus\mathcal E_1$, including its required multiplicities and volume-sector structure, is already determined by $r$. The cylinder lift preserves this feature: it converts the intertwiner parity into commutation or anticommutation with $\cH$, while the physical antiunitary action, combined with the cylinder and auxiliary factors, gives relations that likewise depend only on $r$. We establish these lifts and their signs below, rather than identify the physical and cylinder antiunitaries directly. The result is an exact real graded-algebra isomorphism, not merely a Morita equivalence, for the specified minimal modules at stable auxiliary rank. Concrete matrix sizes can still vary at fixed $r$.

\subsection{The lift and the universal cylinder symmetries}
We determine the real symmetry algebra by listing a complete set of mutually anticommuting generators $e_a$, their squares, and their signs relative to the cylinder Hamiltonian:
\begin{equation}\label{eq:cylinder-generator-data}
 e_ae_b=-e_be_a\quad(a\ne b),\qquad
 e_a^2=\pm1,\qquad e_a\cH=\eta_a\cH e_a.
\end{equation}
Here $\eta_a=+1$ means commuting and $\eta_a=-1$ anticommuting. This $\cH$-grading is \emph{not} the canonical Clifford grading in which every generator is odd. All operators are real-linear; some are complex-antilinear. We retain the minimal physical representations of Eq.~\eqref{eq:physical-clifford-table}, stable auxiliary rank, and symmetry-compatible domains. ``Complete'' concerns constant internal symmetries of the entire allowed $Q$ family, not accidental symmetries of a texture or coordinate transformations.

Take the physical intertwiner spaces $\mathcal E_p$ constructed in Sec.~\ref{sec:boundary-intertwiners}. Their promotion to the cylinder is
\begin{equation}\label{eq:lift}
 \boxed{\begin{gathered}
 G_p(a)=\tau_1^p\otimes a\otimes\one_{\mathrm{aux}},\\
 G_p(a)\cH=(-1)^p\cH G_p(a).
 \end{gathered}}
\end{equation}
For $p=1$, the factor $\tau_1$ supplies the sign for both $\tau_3\partial_t$ and $i\tau_2\Delta R$; the physical intertwiner supplies it for the kinetic term and $i\tau_1\Gamma\Delta S$. Thus these are symmetries of the full cylinder, not just of its kinetic part.

The two universally present even generators are
\begin{equation}\label{eq:universal}
 \begin{gathered}
 I=G_0(i\one),\qquad T=(R_q\otimes u\otimes b)K,\\
 I^2=T^2=-1,\quad IT=-TI,\quad[I,\cH]=[T,\cH]=0,\\
 R_q=\begin{cases}\tau_1,&q=+1,\\\tau_2,&q=-1,\end{cases}\qquad
 q=\begin{cases}\varsigma,&\text{non-chiral},\\1,&\text{chiral}.\end{cases}
 \end{gathered}
\end{equation}
The auxiliary transpose constraints give $T^2=-1$ in every class. The remaining generators depend on the original physical module, whose antiunitary signs are recorded in Table~\ref{tab:input-complete}. Their inherited action is
\begin{equation}\label{eq:crossed-relation}
 \boxed{T G_p(a)T^{-1}=G_p\bigl(q^pua^*u^{-1}\bigr).}
\end{equation}

\begin{table*}[t]\centering\small
\caption{Physical Clifford signs and their antiunitary cylinder lifts, with the kinetic-matrix constraint displayed in every row. The last three columns are physical representation data, not a claim that $uK$ alone is a fixed-$Q$ cylinder symmetry. A dash denotes the absence of an imposed physical chiral operator.}
\label{tab:input-complete}
\begin{tabular}{@{}lcccc@{}}
\toprule
Class&$T$&$uu^*$&Kinetic constraint&Chiral constraint\\\midrule
AI&$(\tau_1\otimes u\otimes J)K$&$+1$&$u\gamma^{\mu *}u^{-1}=-\gamma^\mu$&---\\
AII&$(\tau_1\otimes u)K$&$-1$&$u\gamma^{\mu *}u^{-1}=-\gamma^\mu$&---\\
D&$(\tau_2\otimes u)K$&$+1$&$u\gamma^{\mu *}u^{-1}=+\gamma^\mu$&---\\
C&$(\tau_2\otimes u\otimes J)K$&$-1$&$u\gamma^{\mu *}u^{-1}=+\gamma^\mu$&---\\
BDI&$(\tau_1\otimes u\otimes J)K$&$+1$&$u\gamma^{\mu *}u^{-1}=-\gamma^\mu$&$u\Gamma^*u^{-1}=+\Gamma$\\
CII&$(\tau_1\otimes u)K$&$-1$&$u\gamma^{\mu *}u^{-1}=-\gamma^\mu$&$u\Gamma^*u^{-1}=+\Gamma$\\
DIII&$(\tau_1\otimes u)K$&$-1$&$u\gamma^{\mu *}u^{-1}=-\gamma^\mu$&$u\Gamma^*u^{-1}=-\Gamma$\\
CI&$(\tau_1\otimes u\otimes J)K$&$+1$&$u\gamma^{\mu *}u^{-1}=-\gamma^\mu$&$u\Gamma^*u^{-1}=-\Gamma$\\\bottomrule
\end{tabular}
\end{table*}

\subsection{Lifting the physical generators and their relations}
\label{sec:derive-eight-algebras}
The boundary algebras, multiplicity generators, and volume-sector operators are already fixed in Sec.~\ref{sec:boundary-intertwiners}. Their cylinder representatives are simply
\begin{equation}\label{eq:lifted-physical-seeds}
 \begin{gathered}
 Q_a=G_0(j_a)\quad(r=3,4,5),\qquad
 L=G_0(\ell)\quad(r=2,6),\\
 P=\begin{cases}
 G_1(\omega_m),&m\text{ even},\\
 G_1(a_{\mathrm{ex}}),&r=2,6.
 \end{cases}
 \end{gathered}
\end{equation}
There is no $P$ at $r=0,4$. Equation~\eqref{eq:lift} makes $Q_a,L$ even and $P$ odd. The physical multiplication relations give
\begin{equation}\label{eq:multiplicity-Clifford-seeds}
 \begin{gathered}
 Q_a^2=-1,\quad Q_xQ_y=-Q_z,\\
 [Q_a,I]=[Q_a,T]=[Q_a,P]=[Q_a,\cH]=0,
 \end{gathered}
\end{equation}
where $P$ is included only when present. In the paired-sector cases, $L^2=P^2=1$, $LP=-PL$, $[I,L]=[I,P]=0$, and $TLT^{-1}=-L$.

The only additional sign comes from the cylinder factor in Eq.~\eqref{eq:crossed-relation}. For the physical odd seed $\xi=\omega_m$ or $a_{\mathrm{ex}}$, write $\Theta\xi\Theta^{-1}=\kappa_\xi\xi$. Section~\ref{sec:boundary-intertwiners} gives $\kappa_\xi=\chi$ or $\epsilon\epsilon_0$, respectively, so
\begin{equation}\label{eq:odd-seed-sign}
 \boxed{TPT^{-1}=q\kappa_\xi P=\begin{cases}
 -P,&r=1,2,5,\\
 +P,&r=3,6,7.
 \end{cases}}
\end{equation}
Thus the cylinder relations are obtained by lifting the physical algebra, not by repeating its module decomposition. The fixed auxiliary pairing is essential: the physical $\Theta^2=\epsilon$ becomes $T^2=-1$ after the lift. The boundary algebra and the cylinder algebra must therefore not be identified by merely renaming $\Theta$ as $T$.

The complete Clifford lists are now short. For the quaternionic multiplicity cases $r=3,4,5$ define
\begin{equation}\label{eq:quaternionic-Clifford-basis}
 \begin{gathered}
 (B_1,B_2,B_3,B_4)=(I,T,ITQ_x,ITQ_y),\\
 B_a^2=(-1,-1,+1,+1)_a,\\
 \Omega_B=B_1B_2B_3B_4=ITQ_z.
 \end{gathered}
\end{equation}
The $B_a$ are mutually anticommuting and all commute with $\cH$. Substituting the seed relations gives Table~\ref{tab:physical-restrictions}; each ordered list generates the full real algebra, with no Morita reduction. For each AZ class and dimension, Eq.~\eqref{eq:rdefinition} selects the generator list and its two sign columns. A monomial in the $e_a$ has cylinder sign equal to the product of its $\eta_a$ signs; the spans of positive- and negative-sign monomials define $\mathfrak C^{(r)}_0$ and $\mathfrak C^{(r)}_1$.

\begin{table*}[t]\centering\small
\caption{Complete real graded symmetry algebras for the minimal cylinder family. Within each ordered generator list, distinct generators anticommute; the square and $\eta_a$ tuples follow the same order, with $e_a\cH=\eta_a\cH e_a$. The boundary seeds are derived in Sec.~\ref{sec:boundary-intertwiners} and lifted by Eq.~\eqref{eq:lifted-physical-seeds}. The last two columns give the Clifford and real matrix presentations of the full algebra and its $\cH$-even subalgebra, respectively. The latter consists of positive-cylinder-sign monomials, not canonical even Clifford monomials; $\mathrm{diag}$ denotes the diagonal embedding in the two simple summands. In particular, $r=2,4$ have the same ungraded algebra but different cylinder gradings.}
\label{tab:physical-restrictions}
\setlength{\tabcolsep}{3pt}
\begin{tabular}{@{}clcccc@{}}
\toprule
$r$&Ordered generators $(e_a)$&Squares $(e_a^2)$&Signs $(\eta_a)$&Full algebra $\mathfrak C^{(r)}$&Even subalgebra $\mathfrak C^{(r)}_0$\\\midrule
0&$I,T$&$(-,-)$&$(+,+)$&$\begin{gathered}\Cl_{0,2}\\[-2pt]\cong \HH\end{gathered}$&$\begin{gathered}\Cl_{0,2}\\[-2pt]\cong \HH\end{gathered}$\\
1&$I,T,TP$&$(-,-,+)$&$(+,+,-)$&$\begin{gathered}\Cl_{1,2}\\[-2pt]\cong M_2(\C)\end{gathered}$&$\begin{gathered}\Cl_{0,2}\\[-2pt]\cong \HH\end{gathered}$\\
2&$I,T,TL,TP$&$(-,-,+,+)$&$(+,+,+,-)$&$\begin{gathered}\Cl_{2,2}\\[-2pt]\cong M_4(\R)\end{gathered}$&$\begin{gathered}\Cl_{1,2}\\[-2pt]\cong M_2(\C)\end{gathered}$\\
3&$B_1,B_2,B_3,B_4,\Omega_BP$&$(-,-,+,+,+)$&$(+,+,+,+,-)$&$\begin{gathered}\Cl_{3,2}\\[-2pt]\cong M_4(\R)\oplus M_4(\R)\end{gathered}$&$\begin{gathered}\Cl_{2,2}\\[-2pt]\cong M_4(\R)_{\mathrm{diag}}\end{gathered}$\\
4&$B_1,B_2,B_3,B_4$&$(-,-,+,+)$&$(+,+,+,+)$&$\begin{gathered}\Cl_{2,2}\\[-2pt]\cong M_4(\R)\end{gathered}$&$\begin{gathered}\Cl_{2,2}\\[-2pt]\cong M_4(\R)\end{gathered}$\\
5&$B_1,B_2,B_3,B_4,TQ_zP$&$(-,-,+,+,-)$&$(+,+,+,+,-)$&$\begin{gathered}\Cl_{2,3}\\[-2pt]\cong M_4(\C)\end{gathered}$&$\begin{gathered}\Cl_{2,2}\\[-2pt]\cong M_4(\R)\end{gathered}$\\
6&$I,T,TL,TLP$&$(-,-,+,-)$&$(+,+,+,-)$&$\begin{gathered}\Cl_{1,3}\\[-2pt]\cong M_2(\HH)\end{gathered}$&$\begin{gathered}\Cl_{1,2}\\[-2pt]\cong M_2(\C)\end{gathered}$\\
7&$I,T,ITP$&$(-,-,-)$&$(+,+,-)$&$\begin{gathered}\Cl_{0,3}\\[-2pt]\cong \HH\oplus\HH\end{gathered}$&$\begin{gathered}\Cl_{0,2}\\[-2pt]\cong \HH_{\mathrm{diag}}\end{gathered}$\\\bottomrule
\end{tabular}
\end{table*}

For example, at $r=1$ the seed $P$ commutes with $I$ but anticommutes with $T$; replacing it by $TP$ makes all three listed generators mutually anticommuting. At $r=2$ the two positive-square generators are $TL$ and $TP$, but only $TP$ anticommutes with $\cH$. This illustrates why the square signs and cylinder signs must both be retained. The matrix presentations in the same table follow from the standard real Clifford identities~\cite{ABS}. The real irreducible-module dimension is 4 at $r=0,1,2,3,4,7$ and 8 at $r=5,6$; the split algebras at $r=3,7$ each have two inequivalent irrep types.

For use in the extension problem, $z=IP$ is odd central with $z^2=-1$ at $r=1,5$; $P$ is odd central with $P^2=1$ at $r=3,7$. At $r=2,6$ the grading is $a\mapsto JaJ^{-1}$ with $J=IL$ and $J^2=-1$; at $r=0,4$ every element is even.

For completeness, comparing an arbitrary constant complex-linear symmetry with $\partial_t$ makes it block diagonal (even) or off-diagonal (odd). The full allowed auxiliary mass family removes nontrivial auxiliary action at stable rank; comparing the remaining coefficients gives exactly $G_p(\mathcal E_p)$. Any antilinear symmetry times $T^{-1}$ is complex-linear. The seed lists therefore exhaust the internal algebra; products of the listed $e_a$ supply all its elements. Extra species, nonminimal volume-sector choices, or exceptional auxiliary ranks require a new generator calculation.

\subsection{Why the full graded algebra depends only on \texorpdfstring{$r$}{r}}
\label{sec:r-only-algebra}
We now verify the $r$-only conclusion stated in Eq.~\eqref{eq:r-only-algebra}. The physical-module calculation in Secs.~\ref{sec:dirac-minimality} and \ref{sec:boundary-intertwiners} fixes the available seeds from $r$: multiplicity one at $r=0,1,7$, quaternionic multiplicity two at $r=3,4,5$, and exchanged sectors at $r=2,6$. Its antiunitary relations, combined with the cylinder factor $q$, give Eq.~\eqref{eq:odd-seed-sign}. Since $I,T$ have universal cylinder relations, all lifted generators have identical homogeneous presentations at fixed $r$. Their completeness proves Eq.~\eqref{eq:r-only-algebra}.

Under $D\mapsto D+8$, the physical spinor grows by a factor of 16, but this factor drops out of its intertwiner algebra and the lifted presentation is unchanged. Neither the concrete matrix realization nor the physical spinor dimension is claimed to depend only on $r$.

\subsection{Two-dimensional lifts}
The physical matrices are given in Sec.~\ref{sec:d2-examples}. For the even-$m$ cases, Eq.~\eqref{eq:lifted-physical-seeds} gives $P=G_1(\sigma_z\otimes1_k)$ and, when $k=2$, $Q_a=G_0(1_\sigma\otimes i\rho_a)$. The single-sector DIII and CI models have no odd lift. For CII and BDI, $L=G_0(\rho_z\otimes1)$ and $P=G_1(\rho_x\otimes1)$. Substituting the lifted antiunitary from Table~\ref{tab:input-complete} reproduces the corresponding rows of Table~\ref{tab:physical-restrictions}, including their cylinder grading, without rebuilding the boundary representations.

\section{Quotient indices and stable cylinder zero modes}
\label{sec:quotients}
The general quotient-index formulation for characterizing the stability of zero modes of Hamiltonians with anticommuting symmetries was constructed and systematically investigated by Zheng Zhang and Y.~X.~Zhao in unpublished work~\cite{ZhangZhaoUnpublished}. Here we apply that framework to the elliptic cylinder operators constructed in Sec.~\ref{sec:cylinder}, using their full graded Clifford symmetry algebras derived in Sec.~\ref{sec:full-symmetry}. The task is to determine the quotient index for each cylinder family; its identification with relative texture topology is established in Sec.~\ref{sec:texture-index-isomorphism}.

The purpose of the quotient group $\mathcal I_r$ is to distinguish removable zero modes from those protected by the cylinder symmetry. Under the Fredholm assumptions, the zero-mode space $\mathcal Z=\ker\cH$ is finite-dimensional. Every homogeneous symmetry preserves it: if $v\in\mathcal Z$ and $a\in\mathfrak C_p$, then $\cH(av)=(-1)^pa\cH v=0$. Thus $\mathcal Z$ is a real module of the full symmetry algebra $\mathfrak C^{(r)}$. Whether all its modes can be moved away from zero is not determined by their number alone. At the level of finite-dimensional symmetry-compatible perturbations, it is precisely the question of whether this module extends to the enlarged algebra constructed below.

\subsection{The extension that lifts zero modes}
An invertible symmetry-compatible Hermitian perturbation on $\mathcal Z$ can be spectrally flattened to $h$ with
\begin{equation}\label{eq:extension-rule}
 \boxed{h^\dagger=h,\quad h^2=1,\quad
 ha=(-1)^pah\quad(a\in\mathfrak C_p).}
\end{equation}
It commutes with every even symmetry, including the complex structure $I$, so it is complex-linear; it anticommutes with odd symmetries. Define
\begin{equation}
 \mathfrak C_{\mathrm{ext}}=
 \langle\mathfrak C,h:h^2=1,\ ha=(-1)^{|a|}ah\rangle.
\end{equation}
Conversely, if $\mathcal Z$ extends to a $\mathfrak C_{\mathrm{ext}}$-module, the additional generator supplies an explicit symmetry-compatible lifting perturbation:
\begin{equation}\label{eq:kernel-lifting-perturbation}
 \begin{gathered}
 \delta\cH|_{\mathcal Z}=\varepsilon h,\qquad \varepsilon>0,\\
 (\cH+\delta\cH)|_{\mathcal Z}=\varepsilon h,\qquad
 (\varepsilon h)^2=\varepsilon^2\one.
 \end{gathered}
\end{equation}
Every zero mode is thereby shifted to $+\varepsilon$ or $-\varepsilon$. Complete lifting is possible exactly when $\mathcal Z$ can be decomposed into restrictions of irreducible $\mathfrak C_{\mathrm{ext}}$-modules. Removing such extendible sectors leaves the residual obstruction. Passing to stable equivalence gives the quotient below: a nonzero class of $\mathcal Z$ is the topologically stable residue and certifies zero modes that cannot all be lifted by symmetry-compatible perturbations, although individual modes may pair and move away from zero. Applying the general quotient-index construction~\cite{ZhangZhaoUnpublished}, the obstruction group is
\begin{equation}\label{eq:moduleindex}
 \boxed{\mathcal I_r=
 \frac{\mathscr M(\mathfrak C^{(r)})}
 {\operatorname{im}\bigl[\Res:\mathscr M(\mathfrak C^{(r)}_{\mathrm{ext}})
 \to\mathscr M(\mathfrak C^{(r)})\bigr]}.}
\end{equation}
The protected zero-mode index is the image of $[\mathcal Z]$ in $\mathcal I_r$. Here $\mathscr M$ is the Grothendieck group of ordinary finite-dimensional real modules of the displayed algebras. The numerator therefore contains modules without a chosen module grading, although the algebra retains its $\cH$-grading. This is a module-extension obstruction of the Atiyah--Bott--Shapiro type~\cite{ABS,AtiyahSingerSkew,BCLR}.

\emph{Equivalently, $\mathcal I_r$ is the group of modules modulo graded modules after forgetting their grading.} More precisely, let $\mathscr M_{\mathrm{gr}}(\mathfrak C^{(r)})$ be the Grothendieck group of finite-dimensional real graded modules, with grading-preserving isomorphisms. The forgetful map $\operatorname{For}$ discards the module decomposition; the fixed algebra grading still determines which decompositions are compatible. Then
\begin{equation}\label{eq:moduleindex-forgetful}
 \boxed{\mathcal I_r=
 \frac{\mathscr M(\mathfrak C^{(r)})}
 {\operatorname{im}\bigl[\operatorname{For}:
 \mathscr M_{\mathrm{gr}}(\mathfrak C^{(r)})
 \to\mathscr M(\mathfrak C^{(r)})\bigr]}.}
\end{equation}
The image is essential: a gradable module need not have a unique grading.

Indeed, an action of the extension generator $h$ is precisely a compatible module-grading involution. Its eigenspaces give
\begin{equation}\label{eq:grading-involution}
 \begin{gathered}
 \mathcal Z=\mathcal Z_+\oplus\mathcal Z_-,\qquad
 h|_{\mathcal Z_\sigma}=\sigma\one,\\
 a\mathcal Z_\sigma\subseteq\mathcal Z_{(-1)^p\sigma}
 \quad(a\in\mathfrak C_p,\ \sigma=\pm1).
 \end{gathered}
\end{equation}
Conversely, any such graded module defines an extension action by this $h$; a compatible inner product makes $h$ self-adjoint. Thus restriction from $\mathfrak C_{\mathrm{ext}}$ is the same as forgetting the module grading, and the denominator in Eq.~\eqref{eq:moduleindex} is exactly that in Eq.~\eqref{eq:moduleindex-forgetful}. The grading used here is induced by the cylinder signs in Table~\ref{tab:physical-restrictions}, not canonical all-odd Clifford parity. The quotient measures the stable obstruction to equipping a kernel module with this compatible grading, or equivalently with a symmetry-compatible gap.

The quotient permits all finite-dimensional symmetry-compatible lifting operators. It does \emph{not} assume that every such operator can be realized by a local change of the particular field $Q$. That realization, and the analytic spectral-flow correspondence, are separate from the algebraic calculation.

\begin{table*}[t]\centering\small
\caption{Extension algebras and restriction maps for the grading in Table~\ref{tab:physical-restrictions}. For each AZ class and dimension, $r=s-D-1\pmod8$ selects the corresponding row.}
\label{tab:extensions}
\setlength{\tabcolsep}{4pt}
\begin{tabular}{@{}clllc@{}}
\toprule
$r$&$\mathfrak C^{(r)}$&$\mathfrak C_{\mathrm{ext}}^{(r)}$&$\Res$ on multiplicities&$\mathcal I_r$\\\midrule
0&$\HH$&$\HH\oplus\HH$&$(a,b)\mapsto a+b$&$0$\\
1&$M_2(\C)$&$M_2(\HH)$&$n\mapsto2n$&$\Z_2$\\
2&$M_4(\R)$&$M_4(\C)$&$n\mapsto2n$&$\Z_2$\\
3&$M_4(\R)\oplus M_4(\R)$&$M_8(\R)$&$n\mapsto(n,n)$&$\Z$\\
4&$M_4(\R)$&$M_4(\R)\oplus M_4(\R)$&$(a,b)\mapsto a+b$&$0$\\
5&$M_4(\C)$&$M_8(\R)$&$n\mapsto n$&$0$\\
6&$M_2(\HH)$&$M_4(\C)$&$n\mapsto n$&$0$\\
7&$\HH\oplus\HH$&$M_2(\HH)$&$n\mapsto(n,n)$&$\Z$\\\bottomrule
\end{tabular}
\end{table*}

\subsection{Explicit calculation of the restriction maps}
\paragraph{$r=0,4$: all symmetries are even.}
The generator $h$ is central and its two eigenvalues split the extension into two copies of $\mathfrak C$. Each extension irrep restricts to the original irrep. Hence $\Res(a,b)=a+b$ is surjective and
\begin{equation}
 \mathcal I_0=\mathcal I_4=0.
\end{equation}

\paragraph{$r=1$: quaternionic even algebra.}
Write $\mathfrak C=\HH\oplus z\HH$, $z^2=-1$. The extension adds $h$ commuting with $\HH$ and anticommuting with $z$. Therefore
\begin{equation}
 \mathfrak C_{\mathrm{ext}}\simeq
 \HH\otimes\Cl_{1,1}\simeq\HH\otimes M_2(\R)=M_2(\HH).
\end{equation}
Its real eight-dimensional irrep restricts to two real four-dimensional $M_2(\C)$ irreps. Thus
\begin{equation}
 \boxed{\mathcal I_1=\Z/2\Z,\qquad \nu_1=m\pmod2.}
\end{equation}

\paragraph{$r=2$: an inner grading on real matrices.}
The grading is $a\mapsto JaJ^{-1}$ with $J^2=-1$. Since $h$ commutes with $J$ and implements the same grading, $z=J^{-1}h$ is central and $z^2=-1$. Consequently,
\begin{equation}
 \mathfrak C_{\mathrm{ext}}=M_4(\R)\otimes\C=M_4(\C).
\end{equation}
Its real eight-dimensional irrep restricts to two four-dimensional real irreps, giving
\begin{equation}
 \boxed{\mathcal I_2=\Z/2\Z,\qquad \nu_2=m\pmod2.}
\end{equation}

\paragraph{$r=5,6$: one full irrep is already extendible.}
For $r=5$, the same $z,h$ argument as for $r=1$ starts with even algebra $M_4(\R)$ and gives $M_8(\R)$. For $r=6$, the same inner-grading argument as for $r=2$ gives
\begin{equation}
 M_2(\HH)\otimes\C\simeq M_4(\C).
\end{equation}
In each case both the original irrep and the extension irrep have real dimension eight; restriction is one-to-one on generators. Therefore
\begin{equation}
 \boxed{\mathcal I_5=\mathcal I_6=\Z/\Z=0.}
\end{equation}
This is the extension-theoretic reason the candidate parity does not survive at these entries.

\paragraph{$r=3,7$: paired inequivalent irreps.}
Write $\mathfrak C=A\oplus A$, with the grading exchanging the summands. Adjoining $h$ gives $M_2(A)$. One extension irrep restricts to one irrep of each type, so
\begin{equation}
\boxed{\begin{gathered}
 \mathcal I_{3,7}=\frac{\Z\oplus\Z}{\{(n,n):n\in\Z\}}\simeq\Z,\\
 \Ind=m_+-m_-.
\end{gathered}}
\end{equation}

\subsection{All class--dimension quotients and their meaning}
The result of the algebraic calculation is
\begin{equation}\label{eq:quotient-pattern}
\begin{array}{c|cccccccc}
 r&0&1&2&3&4&5&6&7\\\hline
 \mathcal I_r&0&\Z_2&\Z_2&\Z&0&0&0&\Z.
\end{array}
\end{equation}
Its class--dimension form is Table~\ref{tab:64quotients}.

\begin{table*}[t]\centering\small
\caption{Cylinder module-extension quotient for every boundary dimension $D=0,\ldots,7$. These are algebraic obstruction groups, not the bulk classification table.}
\label{tab:64quotients}
\setlength{\tabcolsep}{5pt}
\begin{tabular}{@{}lcccccccc@{}}
\toprule
Class&$D=0$&$D=1$&$D=2$&$D=3$&$D=4$&$D=5$&$D=6$&$D=7$\\\midrule
AI&$\Z$&$0$&$0$&$0$&$\Z$&$\Z_2$&$\Z_2$&$0$\\
BDI&$0$&$\Z$&$0$&$0$&$0$&$\Z$&$\Z_2$&$\Z_2$\\
D&$\Z_2$&$0$&$\Z$&$0$&$0$&$0$&$\Z$&$\Z_2$\\
DIII&$\Z_2$&$\Z_2$&$0$&$\Z$&$0$&$0$&$0$&$\Z$\\
AII&$\Z$&$\Z_2$&$\Z_2$&$0$&$\Z$&$0$&$0$&$0$\\
CII&$0$&$\Z$&$\Z_2$&$\Z_2$&$0$&$\Z$&$0$&$0$\\
C&$0$&$0$&$\Z$&$\Z_2$&$\Z_2$&$0$&$\Z$&$0$\\
CI&$0$&$0$&$0$&$\Z$&$\Z_2$&$\Z_2$&$0$&$\Z$\\
\bottomrule
\end{tabular}
\end{table*}

The cylinder quotient is distinct from the physical bulk classification $KO^{-r}(\mathrm{pt})$~\cite{Kitaev,RyuReview}. Its pattern is instead
\begin{equation}
 \mathcal I_r\simeq KO^{r-3}(\mathrm{pt}).
\end{equation}
For positive-dimensional textures, this agrees with the stable same-dimensional homotopy of the compact localization target. Using $\mathcal M_s\simeq R_{4-s}$, one obtains~\cite{Bott1959,AtiyahReality,RyuReview}
\begin{equation}\label{eq:target-homotopy-KO}
 \begin{aligned}
 \pi_D(\mathcal M_s)&\cong\pi_0(R_{4-s+D})\\
 &=\pi_0(R_{3-r})\cong KO^{r-3}(\mathrm{pt}).
 \end{aligned}
\end{equation}
Thus, in the stable range and for $D>0$,
\begin{equation}\label{eq:quotient-homotopy-agreement}
 \boxed{\mathcal I_r\cong\pi_D(\mathcal M_s).}
\end{equation}
Agreement of groups alone does not identify the physical texture-to-index map. Section~\ref{sec:texture-index-isomorphism} proves that identification by constructing a texture of unit index in each nonzero case. For $D=0$, fixed saddle rank and component conventions must additionally be specified; an unrestricted stable component group cannot automatically be substituted for a fixed connected Grassmannian.

\section{The physical texture-to-index theorem}
\label{sec:texture-index-isomorphism}
Section~\ref{sec:quotients} has constructed the quotient index that characterizes stable cylinder zero modes. The group isomorphism in Eq.~\eqref{eq:quotient-homotopy-agreement} suggests an index theorem identifying that index with relative texture topology. We now prove this physical identification, not just an agreement of abstract groups. Take $D>0$, stable auxiliary rank, one minimal physical module of Eq.~\eqref{eq:physical-clifford-table}, and smooth based textures constant outside a compact set in $\R^D$. Use their sufficiently slowly varying representatives and the infinite-cylinder Fredholm realization. The index theorem states
\begin{equation}\label{eq:physical-index-isomorphism}
 \boxed{\begin{gathered}
 \Phi_{s,D}:\pi_D(\mathcal M_s)\xrightarrow{\;\cong\;}\mathcal I_r,\\
 \Phi_{s,D}([Q_1]-[Q_0])
 =\Ind_{\mathfrak C}\cH[Q_t],
 \end{gathered}}
\end{equation}
where $\Ind_{\mathfrak C}\cH$ means the kernel-module class in the quotient, not the ordinary complex index. Integer signs require a choice of orientation and irrep labeling.

\begin{figure*}[t]
 \centering
 \resizebox{\textwidth}{!}{
\begin{tikzpicture}[x=1cm,y=1cm,>=stealth,
  every node/.style={font=\small,align=center},
  edge/.style={draw=black!65,line width=.65pt},
  flow/.style={->,draw=black!70,line width=.8pt},
  defect/.style={draw=orange!65!black,fill=orange!9,line width=.9pt},
  note/.style={font=\footnotesize,align=center}]
\definecolor{textureblue}{RGB}{42,91,128}
\definecolor{defectorange}{RGB}{166,83,36}
\path[use as bounding box] (0,-1.65) rectangle (17.4,4.05);

\node[font=\small\bfseries] at (2.25,3.8) {(a) Cylinder interpolation};
\node[font=\small\bfseries] at (8.65,3.8) {(b) Localized obstruction};
\node[font=\small\bfseries] at (15.05,3.8) {(c) Canonical defect};

\fill[textureblue!4] (.95,.65) rectangle (3.55,2.95);
\fill[textureblue!4] (2.25,.65) ellipse (1.3 and .22);
\draw[edge] (.95,.65)--(.95,2.95) (3.55,.65)--(3.55,2.95);
\draw[edge,fill=textureblue!9] (2.25,2.95) ellipse (1.3 and .22);
\draw[edge] (.95,.65) arc[start angle=180,end angle=360,x radius=1.3,y radius=.22];
\draw[edge,dashed] (.95,.65) arc[start angle=180,end angle=0,x radius=1.3,y radius=.22];
\begin{scope}
\clip (.96,.85) rectangle (3.54,2.7);
\foreach \y in {1.05,1.45,1.85,2.25}{
  \draw[textureblue!45,line width=.65pt]
    (.85,\y) .. controls (1.6,\y+.43) and (2.7,\y-.43) .. (3.65,\y);
}
\end{scope}
\node[font=\small] at (2.25,2.95) {$Q_1$};
\node[fill=white,inner sep=1pt] at (2.25,.65) {$Q_0$};
\draw[flow] (.53,.7)--(.53,2.9) node[midway,left] {$t$};
\node[note] at (2.25,-.05) {$Q_t=(1-t)Q_0+tQ_1$};
\node[note] at (2.25,-.48) {Symmetry-compatible self-energy};

\draw[flow] (4.05,1.8)--(6.55,1.8);
\node[note] at (5.3,2.37) {Adiabatic\\deformation};
\node[note] at (5.3,1.22) {Preserve symmetry\\and endpoint gaps};

\fill[textureblue!4] (7.35,.65) rectangle (9.95,2.95);
\fill[textureblue!4] (8.65,.65) ellipse (1.3 and .22);
\draw[edge] (7.35,.65)--(7.35,2.95) (9.95,.65)--(9.95,2.95);
\draw[edge,fill=textureblue!9] (8.65,2.95) ellipse (1.3 and .22);
\draw[edge] (7.35,.65) arc[start angle=180,end angle=360,x radius=1.3,y radius=.22];
\draw[edge,dashed] (7.35,.65) arc[start angle=180,end angle=0,x radius=1.3,y radius=.22];
\node at (8.65,2.95) {$Q_1$};
\node[fill=white,inner sep=1pt] at (8.65,.65) {$Q_0$};
\draw[defect] (8.65,1.8) circle (.8);
\node[text=defectorange] at (8.65,1.8) {$B^{D+1}$};
\node[note,fill=white,inner sep=1pt] at (8.65,-.05) {$\partial B^{D+1}=S^D$};
\node[note] at (8.65,-.48) {Relative class $[Q_1]-[Q_0]$};

\draw[flow] (10.55,1.8)--(13.05,1.8);
\node[note] at (11.8,2.37) {Choose $Q_B(n)$\\on $S^D$};
\node[note] at (11.8,1.22) {Linear filling\\of the ball};

\draw[defect] (15.05,1.8) circle (1.03);
\draw[draw=defectorange!50,dashed] (14.02,1.8)
  arc[start angle=180,end angle=0,x radius=1.03,y radius=.24];
\draw[draw=defectorange!50] (14.02,1.8)
  arc[start angle=180,end angle=360,x radius=1.03,y radius=.24];
\shade[inner color=defectorange!40,outer color=orange!9]
  (15.05,1.8) circle (.48);
\foreach \angle in {30,90,150,210,270,330}{
  \draw[->,draw=defectorange!75,line width=.6pt]
    (15.05,1.8)++(\angle:.91)--++(\angle+180:.52);
}
\fill[defectorange] (15.05,1.8) circle (.055);
\node[note] at (15.05,3.12) {$Q_B(n)$ on $S^D$};
\node[note] at (15.05,.4) {Center: $M(0)=0$};
\node[note] at (15.05,-.05) {$n_a\mapsto y_a$};
\node[note] at (15.05,-.48) {Solve $\ker H_B$ exactly};

\draw[black!20,line width=.45pt] (.35,-.87)--(17.05,-.87);
\node at (8.7,-1.27)
  {$[\ker\widehat H]_{\mathcal I_r}
    =[\ker\widehat H_{\mathrm{def}}]_{\mathcal I_r}
    =[\ker H_B]_{\mathcal I_r}$\quad (one elementary defect)};
\end{tikzpicture}}
 \caption{Geometric idea of the texture--index theorem. (a) Interpolate between the endpoint textures in the symmetry-compatible self-energy space. (b) Deform the field with gapped endpoints so that the obstruction is enclosed in a ball $B^{D+1}$; the normalized field on its boundary $S^D$ carries the relative texture class. Gapped collars and relative-index excision isolate the defect index. (c) For the canonical elementary defect, choose $Q_B(n)$ on $S^D$ and fill the ball linearly, $n_a\mapsto y_a$, giving the solvable operator $H_B$ with a zero of the mass at the center. The index class of the kernel is preserved throughout; the equality shown is for one elementary defect, and general classes are sums of such contributions. Circular outlines schematically represent the ball in arbitrary dimension.}
 \label{fig:texture-index-localization}
\end{figure*}
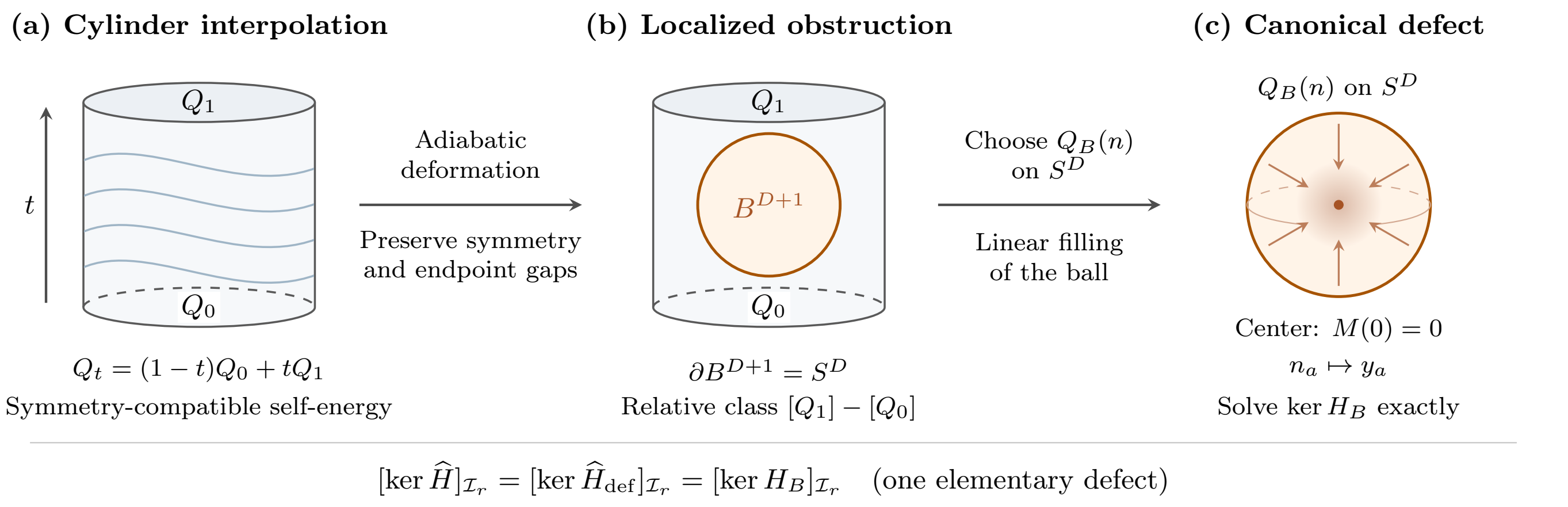

The geometric idea is to deform the cylinder texture until its topological obstruction is concentrated inside a ball, and then compute the index class from a solvable filling of that ball; see Fig.~\ref{fig:texture-index-localization}. Start with $Q_t=(1-t)Q_0+tQ_1$ in the $(D+1)$-dimensional interpolation space $(t,\mathbf x)$. A continuous, symmetry-preserving deformation with gapped asymptotic endpoints leaves the kernel's index class in $\mathcal I_r$ unchanged. The deformation need not keep the interpolating field on the normalized target. It can instead localize the obstruction in a ball $B^{D+1}$, whose normalized boundary texture on $S^D=\partial B^{D+1}$ carries the relative class $[Q_1]-[Q_0]$. Long gapped collars and real relative-index excision isolate this defect contribution without changing the index class~\cite{BunkeRelative,BCLR}.

For the canonical elementary defect, choose the Clifford texture $Q_B(n)$ on the surrounding sphere and fill the ball linearly by replacing the sphere coordinates $n_a$ with Cartesian coordinates $y_a$. The filling leaves the target inside the ball and has vanishing mass at its center. Substituting this field into the cylinder gives the local Dirac operator $H_B$, whose kernel is explicitly solvable. With the gapped completion and excision described below, its kernel class is the index class of the deformed cylinder and hence of the original cylinder. Thus the local calculation evaluates the global invariant because the deformation preserves the index class, not because it preserves every zero mode.

At each nonzero entry, the explicit calculation gives a unit index and establishes that $Q_B$ represents an elementary texture class. General classes are then obtained by combining elementary defects: their index classes add as a signed sum at integer entries or modulo two at $\Z_2$ entries. This order of argument fixes the normalization without assuming in advance that $Q_B$ generates the texture group. The following four subsections establish the gap and localization properties, construct $Q_B$, solve the linear defect, and complete the isomorphism proof.

\subsection{Well-definedness and localization}
\label{sec:index-well-defined}
We first ensure that homotopies of the endpoint textures preserve the asymptotic gap. The gap estimate extends uniformly to the chiral and non-chiral self-energies. For normalized $Q$, both embeddings obey
\begin{equation}\label{eq:normalized-mass-identities}
 \begin{gathered}
 M(Q)M(Q)^\dagger=1,\qquad
 \gamma^\mu M(Q)^\dagger=M(Q)\gamma^\mu,\\
 M(Q)=\begin{cases}
 Q,&Q^\dagger=Q,\quad Q^2=1,\\
 P_+Q+P_-Q^\dagger,&Q^\dagger Q=1.
 \end{cases}
 \end{gathered}
\end{equation}
For the chiral embedding, the second identity uses $\gamma^\mu P_\pm=P_\mp\gamma^\mu$. Direct multiplication in flat space then gives
\begin{equation}\label{eq:allclass-gap-product}
 \begin{aligned}
 \D\D^\dagger&=-\partial^2+\Delta^2
 +\Delta\gamma^\mu\partial_\mu M^\dagger,\\
 \D^\dagger\D&=-\partial^2+\Delta^2
 -\Delta\gamma^\mu\partial_\mu M.
 \end{aligned}
\end{equation}
The first-order derivatives acting on the spinor cancel by Eq.~\eqref{eq:normalized-mass-identities}. Moreover, $\gamma^\mu\partial_\mu M^\dagger=-M(\gamma^\mu\partial_\mu M)M^\dagger$, so the two potential terms have the same norm. Thus, in every real class,
\begin{equation}\label{eq:gap-bound}
 \begin{gathered}
 \D\D^\dagger\geq g_Q\one,\qquad
 \D^\dagger\D\geq g_Q\one,\\
 g_Q=\Delta^2-\Delta\|V_Q\|_\infty,\qquad
 V_Q:=\gamma^\mu\partial_\mu M(Q)^\dagger.
 \end{gathered}
\end{equation}
Hence $\|V_Q\|_\infty<\Delta$ is sufficient for invertibility, without requiring the additional congruence intertwiner $\Xi$. For a non-chiral field this reduces to $\|\gamma^\mu\partial_\mu Q\|_\infty<\Delta$. Compatible compact/covariant realizations require the corresponding operator-domain and gap estimates.

For $D>0$, take a smooth based target homotopy $Q_\lambda$, $0\leq\lambda\leq1$, constant outside a common compact region. The dilation $Q_{\lambda,L}(x)=Q_\lambda(x/L)$ makes the mass gradients uniformly $O(L^{-1})$. For sufficiently large $L$, Eq.~\eqref{eq:gap-bound} therefore keeps both endpoint operators invertible throughout the homotopy. Homotopy invariance of the Fredholm index then identifies their cylinder classes. This construction applies to both mass embeddings; it does not claim a gap for arbitrarily rapidly varying representatives.

Interpolations with the same endpoints are homotopic in the real-linear self-energy space; their asymptotic operators stay fixed and invertible. Hence their Clifford indices agree. Concatenation and real relative-index excision give additivity, including addition of textures in disjoint spatial balls. These properties define $\Phi_{s,D}$. We use the real coefficient version of the relative-index theorem, retaining the full Clifford action; an ordinary complex index would lose precisely the information of interest~\cite{BunkeRelative,BCLR}.

For clarity, the excision step has the following geometric realization. Enclose the varying field in $[t_-,t_+]\times B^D$, with $Q_*$ on the spatial side and $Q_0,Q_1$ on the two temporal faces. After rounding the corners, its normalized boundary mass is a map $S^D\to\mathcal M_s$ representing $[Q_1]-[Q_0]$. Attach long gapped collars and replace the exterior by a radial end carrying the same boundary mass. On that end the squared mass is $\Delta^2$ and its angular derivative is $O(\Delta/R)$, so it is uniformly invertible for large radius. In the cut-and-paste comparison, glue the exterior pieces through such collars: the mass remains normalized and its gradients can be kept uniformly below the gap bound, so this exterior comparison operator is invertible and contributes zero. Real relative-index excision therefore identifies the cylinder class with the local mass-defect class. Changes of the interior extension are compactly supported perturbations and do not change this class. This is the relative-index identification needed below; leaving the normalized target by itself would not establish it~\cite{BunkeRelative}.

\subsection{A texture obeying every physical auxiliary constraint}
\label{sec:index-texture}
Set $d=D+1$. Let $F_a$ be Hermitian, mutually anticommuting auxiliary involutions. Choose them as the kinetic matrices of the auxiliary minimal Dirac module listed in Table~\ref{tab:bott-auxiliary}; this is a construction of the $Q$ matrices, not an extra physical fermion species. For $n\in S^D$, define
\begin{equation}\label{eq:bott-texture}
 Q_B(n)=\begin{cases}
 \displaystyle\sum_{a=0}^{D}n_aF_a,&\text{non-chiral},\\[3pt]
 \displaystyle n_0\one+i\sum_{a=1}^{D}n_aF_a,&\text{chiral}.
 \end{cases}
\end{equation}
The first field is a Hermitian involution and the second is unitary. After a constant unitary change of auxiliary basis, $b$ has the canonical form $1$ or $J$. The constraints in Table~\ref{tab:bott-auxiliary} then give exactly the physical auxiliary conditions of Tables~\ref{tab:nonchiral-input} and \ref{tab:chiral-input}. For $D>0$ the non-chiral field has equal positive and negative ranks. Choose $Q_*=Q_B$ at the base point, make the map constant near that point, and embed it in stable auxiliary rank by adding constant spectator blocks. The reference can be moved within its target component without changing the index. We do not assume that $Q_B$ generates homotopy; its index will prove that fact.

\begin{table*}[t]\centering\small
\caption{Auxiliary Clifford matrices for the explicit texture in Eq.~\eqref{eq:bott-texture}. The auxiliary class specifies the construction of Sec.~\ref{sec:minimal-boundary} in dimension $\ell$; $bK$ is its antiunitary. Its square agrees with the physical auxiliary pairing. In the chiral rows the condition on $F_a$ incorporates complex conjugation of the explicit $i$ in $Q_B$.}
\label{tab:bott-auxiliary}
\setlength{\tabcolsep}{8pt}
\begin{tabular}{@{}lcccc@{}}
\toprule
Physical class&Auxiliary class&$\ell$&$bb^*$&$bF_a^*b^{-1}$\\\midrule
AI&C&$D+1$&$-1$&$+F_a$\\
AII&D&$D+1$&$+1$&$+F_a$\\
D&AI&$D+1$&$+1$&$-F_a$\\
C&AII&$D+1$&$-1$&$-F_a$\\
BDI&C&$D$&$-1$&$+F_a$\\
CII&D&$D$&$+1$&$+F_a$\\
DIII&AI&$D$&$+1$&$-F_a$\\
CI&AII&$D$&$-1$&$-F_a$\\\bottomrule
\end{tabular}
\end{table*}

To illustrate both Tables~\ref{tab:bott-auxiliary} and \ref{tab:bott-index}, consider $D=2$ in the non-chiral class AII and the chiral class CII. Let $\lambda_a$ and $\eta_a$ denote auxiliary Pauli matrices, distinct from the physical spinor matrices of Sec.~\ref{sec:d2-examples}. The displayed fields are finite seed blocks; the based modification and the addition of constant spectator blocks described above are understood. Their small matrix sizes do not replace the stable-rank hypothesis.

\paragraph{Non-chiral example: class AII.}
Here $r=1$, $\dim_\C V_{\min}=2$, and $k=1$. Table~\ref{tab:bott-auxiliary} prescribes an auxiliary class-D module in dimension $\ell=3$, with $bK$ of square $+1$ and $bF_a^*b^{-1}=F_a$. Choose
\begin{equation}\label{eq:d2-AII-auxiliary-texture}
 \begin{gathered}
 F_0=\lambda_x\otimes\one_\eta,\qquad
 F_1=\lambda_z\otimes\eta_x,\\
 F_2=\lambda_z\otimes\eta_z,\qquad b=\one_4,\\
 Q_B^{\mathrm{AII}}(n)=\sum_{a=0}^{2}n_aF_a.
 \end{gathered}
\end{equation}
The three $F_a$ are real symmetric, mutually anticommuting involutions. Consequently, $Q_B^{\mathrm{AII}}$ is real symmetric, Hermitian, and satisfies $Q_B^2=\one_4$ on $S^2$, as required for the AII target. Its positive and negative eigenspaces each have dimension two. The auxiliary spinor has $\dim_\C V_B=4$, so $k_B=4/2=2$. Thus the multiplicity columns of Table~\ref{tab:bott-index} give $(k,k_B)=(1,2)$ at $r=1$; the auxiliary multiplicity is not an extra physical fermion species.

\paragraph{Chiral example: class CII.}
Here $r=2$, and the minimal physical module contains the two volume sectors displayed in Eq.~\eqref{eq:d2-paired-gammas}: $\dim_\C V_{\min}=4$ and $k=2$. Table~\ref{tab:bott-auxiliary} now prescribes an auxiliary class-D module in dimension $\ell=2$. Take
\begin{equation}\label{eq:d2-CII-auxiliary-texture}
 \begin{gathered}
 F_1=\lambda_x,\qquad F_2=\lambda_z,\qquad b=\one_2,\\
 Q_B^{\mathrm{CII}}(n)
 =n_0\one_2+in_1\lambda_x+in_2\lambda_z.
 \end{gathered}
\end{equation}
Again $bb^*=+1$ and $bF_a^*b^{-1}=F_a$. The explicit factors of $i$ make the texture generally non-Hermitian, but it is symmetric and unitary: $Q_B^T=Q_B$ and $Q_B^\dagger Q_B=\one_2$. It therefore obeys the CII constraint, with the chiral mass embedding $M=P_+Q_B+P_-Q_B^\dagger$. Now $\dim_\C V_B=2$ and $k_B=2/2=1$, reproducing $(k,k_B)=(2,1)$ at $r=2$.

To check the remaining columns of Table~\ref{tab:bott-index}, use the oscillator projectors derived in Sec.~\ref{sec:index-local-calculation}. In both examples the local spinor space has complex dimension $2\dim_\C V_{\min}\dim_\C V_B=16$, and the three commuting projectors $(1-\mathsf S_a)/2$, $a=0,1,2$, leave a two-dimensional complex subspace. Thus
\begin{equation}\label{eq:d2-texture-index-examples}
 \begin{array}{c|ccc}
 \text{class}&(k,k_B)&\dim_\C\ker H_B&\Ind_{\mathfrak C}H_B\\\hline
 \mathrm{AII}\ (r=1)&(1,2)&2&1\pmod2\\
 \mathrm{CII}\ (r=2)&(2,1)&2&1\pmod2
 \end{array}
\end{equation}
Both kernels are single real four-dimensional symmetry modules and hence have odd index. The multiplicity two is auxiliary in AII but arises from the symmetry-required physical volume-sector pair in CII.

\subsection{Exact local index and its normalization}
\label{sec:index-local-calculation}
Fill the sphere by replacing $n_a$ with Cartesian coordinates $y_a$ in Eq.~\eqref{eq:bott-texture}. After rescaling positive coefficients, the physical cylinder near this defect is
\begin{equation}\label{eq:bott-oscillator}
 H_B=i\sum_{a=0}^{D}c_a\partial_{y_a}
       +\sum_{a=0}^{D}\mathsf B_a y_a,
\end{equation}
with tensor identities suppressed and
\begin{equation}\label{eq:bott-coefficients}
 \begin{gathered}
 c_0=\tau_3,\qquad c_\mu=\tau_1\gamma^\mu,\\
 \mathsf B_a=-\tau_2F_a\quad\text{(non-chiral)},\\
 \mathsf B_0=-\tau_2,\quad \mathsf B_\mu=-\tau_1\Gamma F_\mu
 \quad\text{(chiral)}.
 \end{gathered}
\end{equation}
All $2d$ matrices are Hermitian involutions and mutually anticommute. These are the actual coefficients of Eqs.~\eqref{eq:cyl-non} and \eqref{eq:cyl-ch}, with the same real symmetry generators. A radial saturation of the linear mass gives a bounded mass of fixed magnitude at infinity; its angular map is $Q_B$. The saturating and linear models have the same defect index by excision across a large sphere with a gapped collar. Thus one may compute the cylinder class with the oscillator rather than assume an index formula for it~\cite{BunkeRelative}.

We now derive the kernel explicitly. Work with the self-adjoint oscillator realization on $L^2(\R^d)\otimes W$, where $d=D+1$ and $W=\C^2\otimes V_{\min}\otimes V_B$ is the internal spinor space; $V_B$ denotes the auxiliary module. The following differential identities hold first on Schwartz spinors and extend to the corresponding operator or quadratic-form domains.

\paragraph{Squaring the defect operator.}
The mixed kinetic--mass terms are
\begin{equation}\label{eq:bott-mixed-terms}
 \begin{aligned}
 &\sum_{a,b=0}^{D}\bigl[(ic_a\partial_{y_a})(\mathsf B_b y_b)
       +(\mathsf B_b y_b)(ic_a\partial_{y_a})\bigr]\\
 &\quad=i\sum_{a=0}^{D}c_a\mathsf B_a
       +i\sum_{a,b=0}^{D}\{c_a,\mathsf B_b\}y_b\partial_{y_a}\\
 &\quad=\sum_{a=0}^{D}\mathsf S_a,\qquad
 \mathsf S_a:=ic_a\mathsf B_a.
 \end{aligned}
\end{equation}
The derivative terms cancel by the Clifford relations. Those relations also eliminate the cross terms within the kinetic and mass squares, giving $H_B^2=-\partial_y^2+|y|^2+\sum_a\mathsf S_a$. Moreover,
\begin{equation}\label{eq:bott-spin-involutions}
 \mathsf S_a^\dagger=\mathsf S_a,\qquad
 \mathsf S_a^2=1,\qquad
 [\mathsf S_a,\mathsf S_b]=0.
\end{equation}
The last identity follows because exchanging two distinct products $c_a\mathsf B_a$ and $c_b\mathsf B_b$ entails four Clifford anticommutations. Thus their simultaneous eigenvalues are signs.

\paragraph{Positivity and the Gaussian zero modes.}
Introduce the oscillator annihilation and creation operators
\begin{equation}\label{eq:bott-ladder-operators}
 \mathfrak a_a=\frac{\partial_{y_a}+y_a}{\sqrt2},\qquad
 \mathfrak a_a^\dagger=\frac{-\partial_{y_a}+y_a}{\sqrt2}.
\end{equation}
Since $-\partial_{y_a}^2+y_a^2=2\mathfrak a_a^\dagger\mathfrak a_a+1$, the square becomes a sum of nonnegative operators:
\begin{equation}\label{eq:bott-positive-square}
 \boxed{H_B^2=\sum_{a=0}^{D}
 \bigl[2\mathfrak a_a^\dagger\mathfrak a_a+(1+\mathsf S_a)\bigr].}
\end{equation}
In particular, for a normalizable zero mode the associated quadratic form gives
\begin{equation}\label{eq:bott-zero-mode-conditions}
 \begin{aligned}
 0=\|H_B\psi\|^2
 &=\sum_{a=0}^{D}\left[2\|\mathfrak a_a\psi\|^2
       +\tfrac12\|(1+\mathsf S_a)\psi\|^2\right],\\
 \mathfrak a_a\psi&=0,\qquad (1+\mathsf S_a)\psi=0
 \quad\text{for every }a.
 \end{aligned}
\end{equation}
Every summand must vanish separately. Solving $(\partial_{y_a}+y_a)\psi=0$ in every coordinate gives $\psi(y)=e^{-|y|^2/2}v$, with $v\in W$ constant. The remaining conditions select $\mathsf S_a v=-v$ for every $a$. Equivalently, each oscillator contributes a zero-point energy $+1$ that is canceled by the internal eigenvalue $-1$; any oscillator excitation raises the energy and cannot be a zero mode.

The sign can also be checked directly in the first-order operator. Since $ic_a\mathsf B_a v=-v$ implies $\mathsf B_a v=ic_a v$, one has
\begin{equation}\label{eq:bott-direct-kernel-check}
 \begin{aligned}
 H_B\bigl(e^{-|y|^2/2}v\bigr)
 &=e^{-|y|^2/2}\sum_{a=0}^{D}y_a(\mathsf B_a-ic_a)v\\
 &=0.
 \end{aligned}
\end{equation}
Thus the conditions are both necessary and sufficient. In particular, the notation $e^{-|y|^2/2}\bigcap_a\ker(\mathsf S_a+1)$ below means the space of wavefunctions $y\mapsto e^{-|y|^2/2}v$ with $v$ in that common internal eigenspace; the Gaussian normalization can be absorbed into $v$.

\paragraph{Counting the internal zero modes.}
The orthogonal projector onto the allowed internal spinors is
\begin{equation}\label{eq:bott-kernel-projector}
 \Pi_{\mathrm{osc}}=\prod_{a=0}^{D}\frac{1-\mathsf S_a}{2}.
\end{equation}
Commutativity alone would not guarantee that every condition halves the dimension. Here the Clifford generators additionally obey
\begin{equation}\label{eq:bott-sign-flip}
 \{c_a,\mathsf S_a\}=0,\qquad
 [c_a,\mathsf S_b]=0\quad(b\ne a).
\end{equation}
The invertible map $c_a$ therefore exchanges the two $\mathsf S_a$ eigenspaces while preserving all other sign conditions. All $2^d$ simultaneous sign sectors have the same dimension, so $\operatorname{rank}_\C\Pi_{\mathrm{osc}}=\dim_\C W/2^d$. Combining the spatial solution and this internal count yields
\begin{equation}\label{eq:bott-kernel}
 \begin{gathered}
 H_B^2=-\partial_y^2+|y|^2+\sum_{a=0}^{D}\mathsf S_a,\\
 \ker H_B=e^{-|y|^2/2}\bigcap_{a=0}^{D}\ker(\mathsf S_a+1),\\
 \dim_\C\ker H_B=\frac{2\dim_\C V_{\min}\dim_\C V_B}{2^{D+1}}
 =k\,k_B.
 \end{gathered}
\end{equation}
Here $k_B=\dim_\C V_B/2^{\lfloor\ell/2\rfloor}$ and $k$ is the physical multiplicity of Table~\ref{tab:physical-multiplicities}. Thus the Gaussian contributes a single spatial state, while all zero-mode degeneracy comes from the common Clifford eigenspace.

The auxiliary construction has residue $r_B=r-3\pmod8$, so Eq.~\eqref{eq:physical-clifford-table} fixes all entries of Table~\ref{tab:bott-index}. In particular, the kernel is one real four-dimensional symmetry module at $r=1,2$, not two; its index is odd. At $r=3,7$, the auxiliary odd Clifford volume is scalar on the chosen simple sector, and $P=\pm\prod_a\mathsf S_a$. The kernel therefore has a single $P$ eigenvalue and consists of one integer-index irrep. Reversing the defect orientation exchanges its sign.

\begin{table}[tb]\centering\small
\caption{Physical local index of the auxiliary texture. The last column is the kernel class in $\mathcal I_r$, not the raw zero-mode dimension. Constant spectator blocks contribute zero.}
\label{tab:bott-index}
\begin{tabular}{@{}ccccc@{}}
\toprule
$r$&$k$&$k_B$&$\dim_\C\ker H_B$&$\Ind_{\mathfrak C}H_B$\\\midrule
0&1&2&2&0\\
1&1&2&2&$1\pmod2$\\
2&2&1&2&$1\pmod2$\\
3&2&1&2&$\pm1$\\
4&2&1&2&0\\
5&2&2&4&0\\
6&2&2&4&0\\
7&1&2&2&$\pm1$\\\bottomrule
\end{tabular}
\end{table}

\subsection{Isomorphism in every case}
\label{sec:index-isomorphism-proof}
Equation~\eqref{eq:quotient-homotopy-agreement} identifies the abstract source and target groups. Having established the additive physical map and evaluated the canonical defect, we can now formulate its sector-detection consequence:
\begin{equation}\label{eq:index-detects-sectors}
 \boxed{\Ind_{\mathfrak C}\cH[Q_t]\ne0
 \quad\Longleftrightarrow\quad[Q_1]\ne[Q_0].}
\end{equation}
This states that the protected cylinder index is nonzero precisely when the endpoints lie in different homotopy sectors. For an additive map between $\Z_2$ groups, detecting the nonzero class suffices to establish an isomorphism. At the integer entries, sector detection gives injectivity but must be supplemented by unit normalization: an injective map $\Z\to\Z$ can multiply a generator by an integer of magnitude greater than one. The explicit unit-index calculation supplies this normalization.

The source and target are each $\Z_2$ for $r=1,2$, each $\Z$ for $r=3,7$, and zero otherwise. The explicitly realized index in Table~\ref{tab:bott-index} makes $\Phi_{s,D}$ surjective; a surjective homomorphism between either of these cyclic groups is an isomorphism. This proves Eq.~\eqref{eq:physical-index-isomorphism}, the criterion above, and the assertion that $Q_B$ is a generator whenever the group is nonzero. Under $D\mapsto D+8$, both spinor dimensions increase by 16 and the oscillator projector denominator increases by $2^8$; the normalized index is unchanged. Thus the construction covers every class and positive dimension, not only the displayed period. It is the physical realization of the Clifford/Bott correspondence, with its normalization fixed rather than inferred from matching tables~\cite{ABS}.

Equation~\eqref{eq:index-detects-sectors} concerns the protected kernel class, not the absence of accidental zero modes when the class vanishes. It does not cover arbitrary rapidly varying representatives, extra physical species, or exceptional auxiliary rank. For $D=0$, the oscillator computation still gives the formal stabilized component index, but a fixed-rank saddle target need not realize that group; no identical claim is made for its individual connected components.

\section{Fixed congruence and the regularized Pfaffian}
\label{sec:congruence-pfaffian}
Sections~\ref{sec:quotients} and \ref{sec:texture-index-isomorphism} have established the stable cylinder index and its identification with relative texture topology. To connect that index to the fermion integral, we first need a regularized Pfaffian whose relative sign is well defined. This section constructs such a representative at every $r=1,2$ entry, including the chiral classes. A single field-independent unitary congruence transforms the original complex antisymmetric kernel into a Hermitian antisymmetric operator; multiplication by $-i$ then gives a real skew-adjoint operator. At a common finite cutoff, its Pfaffian differs from the original one only by a field-independent factor, which cancels in relative phases. The congruence supplies the real structure, while a common symmetry-compatible regulator and reference orientation define the relative regularized Pfaffian. Section~\ref{sec:paired-spectral-flow} then uses Witten's spectral-flow argument to identify this relative sign with the index and derive the theta weight.

\subsection{Complex Pfaffians and the relative phase}
\label{sec:relative-Pf-phase}
With the common regulator and fermionic measure of Sec.~\ref{sec:fermionic-action}, define the relative phase for nonzero endpoint Pfaffians by
\begin{equation}\label{eq:relative-phase}
 \mathcal P[Q_1,Q_0]=\frac{F[Q_1]}{|F[Q_1]|}
 \left(\frac{F[Q_0]}{|F[Q_0]|}\right)^{-1}.
\end{equation}
Skew-symmetry does not require reality. At a finite cutoff the Pfaffian is a polynomial in the entries of the complex antisymmetric matrix, with its overall sign fixed by the ordered Grassmann measure. Its nonzero phase and its local transport are
\begin{equation}
 \begin{aligned}
 e^{i\phi[Q]}&=F[Q]/|F[Q]|,\\
 \dot\phi&=\tfrac12\operatorname{Im}\Tr_{\mathrm{reg}}
 (\A_t^{-1}\dot\A_t).
 \end{aligned}
\end{equation}
Complex skew eigenvalues occur in opposite pairs, but their unordered endpoint set fixes only $F^2$, not the Pfaffian orientation. For instance, $\A_t=e^{i\pi t}J_1$ has $F(t)=e^{i\pi t}$ and no zero on $0\leq t\leq1$, despite opposite endpoint phases. Thus an arbitrary complex Pfaffian sign comparison cannot be replaced by zero-crossing parity.

The anomaly spectral prescription instead identifies a suitable paired real spectrum, selects one branch from each pair at a reference configuration, and transports the choice continuously~\cite{WittenSU2,RyuAnomaly}. The following subsections prove an exact fixed-congruence relation for all $r=1,2$ minimal families, including the chiral ones, and states the required regularization. Its original Pfaffian phase can therefore be computed directly from the Hermitian spectral representative; Sec.~\ref{sec:paired-spectral-flow} then connects the branch crossings to our cylinder. Equality of the eigenvalues of the original complex kernel and this representative is neither assumed nor required.

\subsection{The adjoint identity for non-chiral and chiral kernels}
\label{sec:kinetic-grading}
Use the same sign $q$ as in Eq.~\eqref{eq:crossed-relation}: $q=\varsigma$ in non-chiral classes and $q=1$ in chiral classes. Their transpose identities have the uniform form
\begin{equation}\label{eq:uniform-transpose}
 \begin{gathered}
 \B^T=-q\B,\qquad \B\B^*=-q,\\
 \B\D[Q]^{\mathsf t}\B^{-1}=q\D[Q],\qquad
 \B\D[Q]^*\B^{-1}=q\D[Q]^\dagger.
 \end{gathered}
\end{equation}
The last identity follows by taking the adjoint of the preceding one. All identities concern the full operator on a compatible domain.

Seek a constant physical Hermitian involution $\Xi$ in the odd intertwiner space of Eq.~\eqref{eq:graded-commutant}:
\begin{equation}\label{eq:Xi-conditions}
 \boxed{\begin{gathered}
 \Xi^\dagger=\Xi,\qquad\Xi^2=1,\qquad
 \{\Xi,E_j\}=0\quad\text{for every }j,\\
 [\Xi,Q]=[\Xi,Q^\dagger]=0.
 \end{gathered}}
\end{equation}
The last line holds because physical and auxiliary matrices act on separate factors. This intertwiner of the specified self-energy family is not an additional physical AZ symmetry imposed on arbitrary disorder. For a non-chiral kernel, $E_j=\gamma^j$ and $Q^\dagger=Q$, so these conditions give
\begin{equation}\label{eq:gamma-adjoint}
 \boxed{\D[Q]^\dagger=\Xi\D[Q]\Xi.}
\end{equation}
In a chiral class the same odd-intertwiner condition also imposes $\{\Xi,\Gamma\}=0$. To check the self-energy explicitly, write
\begin{equation}\label{eq:chiral-adjoint-check}
 \begin{aligned}
 R&=\tfrac12(Q+Q^\dagger),\quad
 S=\tfrac1{2i}(Q-Q^\dagger),\\
 M(Q)&=R+i\Gamma S,\\
 \Xi\D[Q]\Xi
 &=-\slp+\Delta(R-i\Gamma S)\\
 &=\D[Q]^\dagger.
 \end{aligned}
\end{equation}
Neither $Q^2=1$ nor $Q^\dagger Q=1$ is needed for this adjoint identity; it holds throughout the real-linear interpolation space. The chiral self-energy therefore poses no obstruction when the physical odd intertwiner anticommutes with $\Gamma$ as well as with all kinetic matrices.

Consequently,
\begin{equation}\label{eq:spectral-H}
 \boxed{H^{\mathrm{sp}}[Q]=\Xi\D[Q],\qquad
 (H^{\mathrm{sp}}[Q])^\dagger=H^{\mathrm{sp}}[Q].}
\end{equation}
This is an auxiliary spectral representative, not the physical Hamiltonian $H_0$. To determine its reality structure, let
\begin{equation}\label{eq:Xi-parity}
 u\Xi^*u^{-1}=\kappa_\Xi\Xi,\qquad
 \delta=q\kappa_\Xi,
\end{equation}
and define $\mathcal C_{\mathrm{sp}}=\Xi\B K$. Equation~\eqref{eq:uniform-transpose} gives
\begin{equation}\label{eq:spectral-C}
 \boxed{\begin{aligned}
 \mathcal C_{\mathrm{sp}}^2&=-\delta,\\
 \mathcal C_{\mathrm{sp}}H^{\mathrm{sp}}
 \mathcal C_{\mathrm{sp}}^{-1}&=\delta H^{\mathrm{sp}}.
 \end{aligned}}
\end{equation}
For $\delta=-1$ this is a real structure pairing the Hermitian eigenvalues as $\{-\lambda,+\lambda\}$. For $\delta=+1$ it is a commuting Kramers symmetry; this particular construction does not then give the required real opposite pairs.

\subsection{Existence at every \texorpdfstring{$\Z_2$}{Z2} entry from the physical module}
\label{sec:allclass-congruence}
Section~\ref{sec:boundary-intertwiners} supplies the physical intertwiner $\Xi$ and its antiunitary action; Sec.~\ref{sec:full-symmetry} gives the inherited cylinder sign. Recall that $m=D$ for non-chiral classes and $m=D+1$ for chiral classes.

At $r=1$, $m$ is even and the restricted physical module is a single complex spinor. Choose its normalized volume matrix,
\begin{equation}\label{eq:even-Xi-allclass}
 \Xi=\omega_m=(-i)^{m/2}E_1\cdots E_m.
\end{equation}
It is a Hermitian involution anticommuting with every $E_j$. The inherited seed sign in Eq.~\eqref{eq:odd-seed-sign} is $TPT^{-1}=-P$ for $P=G_1(\omega_m)$. By Eq.~\eqref{eq:crossed-relation}, this is exactly $q\kappa_\Xi=-1$.

At $r=2$, $m$ is odd and the physical antiunitary requires both volume sectors. In the representation of Eq.~\eqref{eq:physical-sector-exchange}, take
\begin{equation}\label{eq:allclass-sector-rep}
 \boxed{E_j=\rho_z\otimes e_j,\qquad
 \Xi=\rho_x\otimes1.}
\end{equation}
This is again a Hermitian odd involution, and Eq.~\eqref{eq:odd-seed-sign} gives $q\kappa_\Xi=-1$. No extra physical species has been added: the two sectors belong to the original minimal real module.

Thus every $\Z_2$ entry of Table~\ref{tab:64quotients} satisfies
\begin{equation}\label{eq:allclass-congruence-criterion}
 \boxed{r=1,2\quad\Longrightarrow\quad
 \D^\dagger=\Xi\D\Xi,\qquad\delta=-1}
\end{equation}
for the specified family. Table~\ref{tab:z2-congruence} lists all entries in one period. In chiral classes it is the parity of $m=D+1$, not of $D$, that selects the volume or exchange construction. At $D=0$ the same finite-dimensional algebra applies: use the empty volume $\omega_0=1$ in class D, and the two $\Gamma$ sectors in DIII. Interpretation of the allowed zero-dimensional saddle components remains a separate question.

\begin{table*}[t]\centering\small
\caption{Fixed-congruence construction at all $\Z_2$ positions for $0\leq D\leq7$. At $r=1$, $m$ is even and $\Xi=\omega_m$; at $r=2$, $m$ is odd and $\Xi=\rho_x\otimes1$ exchanges the symmetry-required volume sectors. Every listed case has $\delta=q\kappa_\Xi=-1$. The $D=0$ cases are algebraic statements, not identifications of fixed-rank component groups.}
\label{tab:z2-congruence}
\begin{tabular}{@{}lccc@{}}
\toprule
Class&Number of physical generators&$D$ at $r=1$&$D$ at $r=2$\\\midrule
AI&$m=D$&6&5\\
BDI&$m=D+1$&7&6\\
D&$m=D$&0&7\\
DIII&$m=D+1$&1&0\\
AII&$m=D$&2&1\\
CII&$m=D+1$&3&2\\
C&$m=D$&4&3\\
CI&$m=D+1$&5&4\\\bottomrule
\end{tabular}
\end{table*}

The converse is false. At $r=5$ the even-$m$ volume also has $q\kappa_\Xi=-1$, but the physical multiplicity is quaternionic and the full quotient is zero. For example, $D=2$ in AI admits the same volume construction as AII, with an additional physical multiplicity. Existence of a Hermitian congruence is therefore not equivalent to a nontrivial $\Z_2$ obstruction. Nor do we infer this existence merely from an abstract quotient group: it follows from the specific physical module and self-energy above. We do not claim an exhaustive nonexistence result at the other entries for every possible choice of intertwiner or representation.

\subsection{Fixed unitary congruence and the finite-cutoff Pfaffian}
\label{sec:fixed-congruence}
In the paired case $\delta=-1$, set
\begin{equation}
 \mathcal S=\Xi\B,\qquad
 \mathcal C_{\mathrm{sp}}=\mathcal S K,\qquad
 \A[Q]=\mathcal S^{-1}H^{\mathrm{sp}}[Q].
\end{equation}
The matrix $\mathcal S$ is unitary and
\begin{equation}\label{eq:S-symmetric}
 \mathcal S^T=-q\kappa_\Xi\mathcal S=\mathcal S,
 \qquad\mathcal S\mathcal S^*=1.
\end{equation}
Its symmetric-unitary Takagi factorization~\cite{Takagi1924} can be chosen once for the entire family:
\begin{equation}\label{eq:S-takagi}
 \mathcal S=UU^T,\qquad U^\dagger U=1.
\end{equation}
Since $\Xi$ and $\B$ are constant, $U$ is independent of $Q$ and the interpolation parameter $t$. Direct substitution gives
\begin{equation}\label{eq:fixed-congruence}
 \boxed{\begin{aligned}
 U^T\A_tU
 &=U^T\mathcal S^{-1}H_t^{\mathrm{sp}}U\\
 &=U^\dagger H_t^{\mathrm{sp}}U=:H'_t.
 \end{aligned}}
\end{equation}
Thus $H'_t$ is \emph{congruent} to the original antisymmetric kernel but \emph{unitarily similar} to the Hermitian spectral representative. Equality of the spectra of $\A_t$ and $H_t^{\mathrm{sp}}$ is neither asserted nor needed. Their kernels are naturally identified.

In this basis $U^\dagger\mathcal C_{\mathrm{sp}}U=K$. Hence
\begin{equation}\label{eq:aux-real}
 \begin{gathered}
 (H'_t)^*=-H'_t,\qquad (H'_t)^{\mathsf t}=-H'_t,\\
 A_{R,t}:=-iH'_t,\qquad
 A_{R,t}^*=A_{R,t},\quad A_{R,t}^{\mathsf t}=-A_{R,t}.
 \end{gathered}
\end{equation}
A Hermitian antisymmetric operator is purely imaginary; $A_{R,t}$ is therefore real skew-adjoint. The Hermitian congruence does not make the fermion kernel real symmetric.

At a common finite cutoff of total dimension $2N_\Lambda$, let $U_\Lambda$ denote the induced fixed change of variables. The Pfaffian congruence law gives the exact polynomial identity
\begin{equation}\label{eq:Pf-congruence}
 \boxed{\begin{aligned}
 F_\Lambda[Q]&:=\Pf\A_\Lambda[Q]\\
 &=(\det U_\Lambda)^{-1}\Pf H'_\Lambda[Q]\\
 &=c_\Lambda\Pf A_{R,\Lambda}[Q],\\
 c_\Lambda&=\frac{i^{N_\Lambda}}{\det U_\Lambda}.
 \end{aligned}}
\end{equation}
The factor $c_\Lambda$ is $Q$ independent. Squaring this identity yields
\begin{equation}\label{eq:Pf-det-sp}
 \begin{gathered}
 F_\Lambda[Q]^2=d_\Lambda\det H^{\mathrm{sp}}_\Lambda[Q],\\
 d_\Lambda=\det(\B_\Lambda^{-1}\Xi_\Lambda),\qquad
 c_\Lambda^2=(-1)^{N_\Lambda}d_\Lambda.
 \end{gathered}
\end{equation}
Unlike the determinant relation alone, Eq.~\eqref{eq:Pf-congruence} retains the sign fixed by the ordered Grassmann measure. Choosing a separate $Q$-dependent congruence for each configuration would require its Jacobian and orientation transport; it would not establish this fixed-family identity.

\subsection{Checks in non-chiral and chiral representatives}
\label{sec:congruence-examples}
For $D=2$ in AII, $\Xi=\sigma_z$, $u=i\sigma_y$, and $b=1$ give
\begin{equation}\label{eq:AII2-spectral}
 H_t^{\mathrm{sp}}=-\sigma_y p_x+\sigma_xp_y
 +\Delta\sigma_zQ_t.
\end{equation}
A constant spin rotation gives the Hermitian representative used in the AII spectral analysis~\cite{RyuAnomaly}. For $D=1$ in AII, the original time-reversal representation already contains two kinetic sectors:
\begin{equation}\label{eq:AII1-congruence}
 \begin{gathered}
 \gamma^1=\rho_z,\quad u=i\rho_y,\quad b=1,\quad\Xi=\rho_x,\\
 H_t^{\mathrm{sp}}=-i\rho_y\partial_x+\Delta\rho_xQ_t,\\
 \mathcal S=-\rho_z,\qquad U=\operatorname{diag}(i,1).
 \end{gathered}
\end{equation}
This explicitly realizes the fixed congruence without an artificial species doubling.

For the chiral example $D=2$ in CII, use the sector-first matrices of Sec.~\ref{sec:d2-examples}. Then
\begin{equation}\label{eq:CII2-congruence}
 \begin{gathered}
 \gamma^{1,2}=\rho_z\otimes\sigma_{x,y},\quad
 \Gamma=\rho_z\otimes\sigma_z,\\
 u=i\rho_y\otimes\sigma_x,\quad b=1,\quad
 \Xi=\rho_x\otimes1,\\
 \mathcal S=-\rho_z\otimes\sigma_x,\qquad
 \mathcal S^T=\mathcal S,\quad\mathcal S^2=1.
 \end{gathered}
\end{equation}
Both $\{\Xi,\gamma^\mu\}=0$ and $\{\Xi,\Gamma\}=0$ hold. With $\Pi_\pm^{\mathcal S}=(1\pm\mathcal S)/2$, the constant matrix $U=\Pi_+^{\mathcal S}+i\Pi_-^{\mathcal S}$ obeys $UU^T=\mathcal S$ and makes Eq.~\eqref{eq:fixed-congruence} explicit for a general allowed chiral $Q$. These examples check existence of the congruence, not the index on a generating texture.

\subsection{Common regularization and the reference orientation}
\label{sec:regularized-real-Pf}
The infinite-dimensional Pfaffian is not an unregulated product. Choose a single regulator for the family, an orientation of the regulated Grassmann measure, and an invertible reference $Q_*$. The cutoff spaces and domains must preserve the fixed change of variables, transpose structure, and inherited real symmetries, or their measure contributions must be included explicitly. At finite cutoff,
\begin{equation}\label{eq:regulated-Pf-ratio}
 \frac{F_\Lambda[Q]}{F_\Lambda[Q_*]}
 =\frac{\Pf A_{R,\Lambda}[Q]}{\Pf A_{R,\Lambda}[Q_*]}.
\end{equation}
Thus $c_\Lambda$ cancels before any continuum limit, even if it has no meaningful standalone limit.

If magnitude renormalization can be chosen with real counterterms $S_{\mathrm{ct},\Lambda}$ compatible with this real structure, the relative regularized Pfaffian can be written
\begin{equation}\label{eq:regularized-Pf-definition}
 \begin{aligned}
 \frac{F[Q]}{F[Q_*]}
 &=\lim_{\Lambda\to\infty}
 e^{-\Delta S_{\mathrm{ct},\Lambda}[Q]}
 \frac{\Pf A_{R,\Lambda}[Q]}{\Pf A_{R,\Lambda}[Q_*]},\\
 \Delta S_{\mathrm{ct},\Lambda}[Q]
 &:=S_{\mathrm{ct},\Lambda}[Q]-S_{\mathrm{ct},\Lambda}[Q_*].
 \end{aligned}
\end{equation}
The positive renormalization factor cannot change the sign. This formula states the regulator matching needed for the relative phase; it is not a proof that every microscopic regularization has these properties. An additional ultraviolet phase or topological counterterm must be matched to the microscopic model rather than silently discarded. Subject to this matching and nonzero endpoint Pfaffians,
\begin{equation}\label{eq:relative-real-pf}
 \boxed{\mathcal P[Q_1,Q_0]
 =\lim_{\Lambda\to\infty}
 \frac{\operatorname{sgn}\Pf A_{R,\Lambda}[Q_1]}
 {\operatorname{sgn}\Pf A_{R,\Lambda}[Q_0]}.}
\end{equation}
Here the limit denotes the stabilized relative sign in the chosen prescription.

A regulator and ordered measure define the finite-cutoff Pfaffian as a polynomial, including its sign. The apparent two-valuedness arises if one instead retains only its determinant or its absolute value. The overall reference orientation is conventional and cancels in a relative sign; selecting unrelated signs in different texture sectors is not part of this fixed definition. More generally the fermionic integral is a section of a Pfaffian line, and a proposed global trivialization must be checked against any identifications of backgrounds~\cite{DaiFreedOriginal,Witten,DaiFreed}. The next section explains the spectral transport of this orientation and the precise connection with Witten's argument.

\section{Witten's spectral flow and the \texorpdfstring{$\Z_2$}{Z2} theta term}
\label{sec:paired-spectral-flow}
Section~\ref{sec:congruence-pfaffian} identifies the original complex Pfaffian with a real skew-adjoint Pfaffian up to a field-independent factor and specifies the required regularization. We now address a different question: how does its sign change between two invertible backgrounds, and how is that change measured by the cylinder index? Witten's argument uses continuous spectral transport rather than an independent choice of square root at each endpoint~\cite{WittenSU2,Witten}. For our families this transport applies to the Hermitian representative already proved congruent to the original kernel. The remaining analytic step identifies its sign transport with the normalized cylinder parity. The index theorem of Sec.~\ref{sec:texture-index-isomorphism} then identifies that parity with the relative homotopy class, so the $\Z_2$ theta weight follows without a new gap proof or generator calculation.

\subsection{What the sign ambiguity means}
\label{sec:Witten-sign-ambiguity}
For a real skew matrix with an oriented two-dimensional block $aJ_1$,
\begin{equation}\label{eq:elementary-Pf-sign}
 \Pf(aJ_1)=a,\qquad \det(aJ_1)=a^2.
\end{equation}
Its determinant or its positive square root $|a|$ loses the sign. Fixing the orientation and following $a$ through zero retains it. This elementary distinction survives regularization: a determinant does not specify the orientation of the Pfaffian line~\cite{Witten}.

In Witten's $SU(2)$ anomaly, transporting the fermionic square root along a path to a gauge-equivalent configuration can reverse its sign. An odd spectral-flow parity then obstructs a continuous gauge-invariant choice on the gauge-identified space of backgrounds; the higher-dimensional mod-two index detects that obstruction~\cite{WittenSU2,Witten}. This is not an arbitrary freedom to redefine the sign separately at every configuration after fixing a fermionic measure.

Our comparison of sigma textures uses the same orientation-flow mechanism but a different endpoint interpretation. The linear interpolation passes through the larger symmetry-compatible self-energy space; distinct textures are not automatically gauge-equivalent. At a common finite cutoff in this affine space, $F_\Lambda[Q]$ is a single-valued polynomial. Witten's transport prescription determines its relative signs from a reference; it does not imply a gauge anomaly without an additional physical identification of endpoints. The two-dimensional AII calculation realizes this mechanism as a relative Pfaffian sign between sigma-model sectors~\cite{RyuAnomaly}.

\subsection{Selected spectral branches and regulated sign transport}
\label{sec:selected-branches}
The real skew matrix $A_{R,\Lambda,t}$ has local orthogonal blocks $a_j(t)J_1$; the Hermitian blocks $ia_j(t)J_1$ have opposite eigenvalues $\pm a_j(t)$. Choose the orientation and one branch from each pair consistently at the reference configuration, and transport the selection continuously. With that convention Eq.~\eqref{eq:Pf-congruence} gives
\begin{equation}\label{eq:branch-Pf}
 F_\Lambda(t)=c_\Lambda\prod_{j=1}^{N_\Lambda}
 \widetilde\lambda_j(t).
\end{equation}
The choice is the continuous lift of the determinant square root, not a principal square root reevaluated at each $t$. The cutoff-dependent magnitude and the factor $c_\Lambda$ are handled by Sec.~\ref{sec:regularized-real-Pf}; the relative sign is transported by the selected branches.

Near an elementary transverse crossing,
\begin{equation}
 \widetilde\lambda(t)=v(t-t_*)+O((t-t_*)^2),\qquad v\ne0,
\end{equation}
one selected factor changes sign. Degenerate crossings must be resolved or counted with their full symmetry multiplicities. For invertible endpoints, the stabilized parity yields
\begin{equation}\label{eq:pair-flow-sign}
 \boxed{\mathcal P[Q_1,Q_0]=(-1)^{N_{\mathrm{pair}}}.}
\end{equation}
Here $N_{\mathrm{pair}}$ counts independent sign-changing pairs along the path modulo two, in the common regulator. The unordered endpoint spectrum fixes only the square; choosing positive eigenvalues afresh would erase the sign. Ordinary signed spectral flow of the full Hermitian spectrum also vanishes because opposite branches cancel. The relevant invariant is instead the real orientation flow of $A_{R,t}$, denoted $\SF_2\{A_{R,t}\}$~\cite{CLPS,BCLR}. For the families of Sec.~\ref{sec:allclass-congruence}, this is a prescription for the original regularized Pfaffian, not a newly defined fermion integral.

\subsection{Unitary reduction of the constructed cylinder}
Introduce a positive stretching parameter $\varepsilon$ and keep the endpoint operators invertible:
\begin{equation}
 \widehat{\D}_\varepsilon=
 \begin{bmatrix}\varepsilon\partial_t&\D_t\\
 -\D_t^\dagger&-\varepsilon\partial_t\end{bmatrix},
 \qquad \widehat H_\varepsilon=i\widehat{\D}_\varepsilon.
\end{equation}
With $W_\Xi=\operatorname{diag}(\Xi,1)$, Eq.~\eqref{eq:gamma-adjoint} implies
\begin{equation}\label{eq:cyl-sp-equivalence}
 \begin{aligned}
 & W_\Xi^\dagger\widehat{\D}_\varepsilon W_\Xi
 =K_\varepsilon,\\
&  K_\varepsilon =
 \begin{bmatrix}\varepsilon\partial_t&H_t^{\mathrm{sp}}\\
 -H_t^{\mathrm{sp}}&-\varepsilon\partial_t\end{bmatrix}=\tau_3\varepsilon\partial_t+i\tau_2H_t^{\mathrm{sp}}.
 \end{aligned}
\end{equation}
In particular,
\begin{equation}\label{eq:cylinder-mode-equation}
 \boxed{K_\varepsilon\Psi=0\ \Longleftrightarrow\
 (\varepsilon\partial_t+\tau_1H_t^{\mathrm{sp}})\Psi=0.}
\end{equation}
This reduction uses only the adjoint identity, so it applies equally to the chiral and non-chiral families in Sec.~\ref{sec:allclass-congruence}. It connects the selected spectral family to the original cylinder, rather than to a different auxiliary operator with an unspecified phase relation.

\subsection{Localized modes at an elementary pair crossing}
Near an isolated crossing, choose $H_t^{\mathrm{sp}}\varphi_t=\lambda(t)\varphi_t$ and $\tau_1\chi_\sigma=\sigma\chi_\sigma$. Ignoring the moving-basis connection at leading adiabatic order, $\Psi=f(t)\chi_\sigma\otimes\varphi_t$ obeys
\begin{equation}
 (\varepsilon\partial_t+\sigma\lambda(t))f(t)=0,
\end{equation}
so
\begin{equation}\label{eq:crossing-gaussian}
 \begin{aligned}
 f(t)&=f(t_*)\exp\!\left[-\frac{\sigma}{\varepsilon}
 \int_{t_*}^{t}\lambda(u)\dd u\right],\\
 f(t)&\sim\exp\!\left[-\frac{\sigma v}{2\varepsilon}
 (t-t_*)^2\right].
 \end{aligned}
\end{equation}
Normalizability selects $\sigma=\operatorname{sgn}v$. The partner branch $-\lambda(t)$ supplies one mode in the opposite $\tau_1$ sector. An elementary pair crossing therefore contributes two complex cylinder modes in this local model.

The oscillator mechanism is also visible directly:
\begin{equation}\label{eq:susp-square}
 \boxed{-K_\varepsilon^2=-\varepsilon^2\partial_t^2
 +(H_t^{\mathrm{sp}})^2-\varepsilon\tau_1\dot H_t^{\mathrm{sp}}.}
\end{equation}
On the linearized branch this becomes
\begin{equation}
 -\varepsilon^2\partial_t^2+v^2(t-t_*)^2
 -\varepsilon\sigma v.
\end{equation}
For $\sigma v>0$, the lowest oscillator eigenvalue is canceled by the last term. A varying eigenbasis contributes connection terms, and modes at distinct crossings can hybridize. Hence the exact global kernel dimension need not equal twice the number of crossings. The protected equality is only modulo the removable modules, as made precise by spectral-flow index theory~\cite{BCLR}.

\subsection{Reduction to a real suspension and the mod-two theorem}
Use the same fixed $U$ and real skew-adjoint operator $A_{R,t}=-iU^\dagger H_t^{\mathrm{sp}}U$ constructed in Sec.~\ref{sec:fixed-congruence}. Equation~\eqref{eq:Pf-congruence} already identifies its Pfaffian phase with that of the original complex $\A_t$. The remaining step is to identify their spectral-flow index with the kernel of our cylinder.

Let $V=\operatorname{diag}(1,-i)$ act in cylinder space. After applying $\operatorname{diag}(U,U)$ to $K_\varepsilon$, a further unitary transformation by $V$ gives
\begin{equation}\label{eq:real-susp}
 L_R=
 \begin{bmatrix}\varepsilon\partial_t&A_{R,t}\\
 A_{R,t}&-\varepsilon\partial_t\end{bmatrix}.
\end{equation}
It is real skew-adjoint and anticommutes with
\begin{equation}
 \mathcal J=\begin{bmatrix}0&1\\-1&0\end{bmatrix},\qquad
 \mathcal J^2=-1.
\end{equation}
The fixed-congruence identity \eqref{eq:Pf-congruence} and the cylinder unitary transformations identify both the original Pfaffian phase and the cylinder spectrum with this real suspension. Thus the real orientation flow computes the original relative phase, even though $\A_t$ need not be entrywise real.

Use a compatible compact spatial realization with a common Sobolev domain and compact resolvent, smooth bounded zeroth-order variation, and invertible asymptotic limits; take the path constant outside a compact $t$ interval. More general infrared realizations must supply the corresponding Fredholm hypotheses. All fixed transformations must preserve the domains, and the regularization must be matched as in Sec.~\ref{sec:regularized-real-Pf}. For $r=1,2$, each irreducible cylinder symmetry module has real dimension four. Writing $\ker\widehat H_\varepsilon=W^{\oplus m}$, the quotient index constructed in Sec.~\ref{sec:quotients} is therefore $\nu_{\mathrm{cyl}}=m\pmod2$, equivalently $\nu_{\mathrm{cyl}}=\frac12\dim_\C\ker\widehat H_\varepsilon\pmod2$. This counts protected module multiplicities, not the parity of the raw complex kernel dimension. Under the assumptions above, the real Clifford-valued spectral-flow--index theorem identifies this same index with the Pfaffian orientation flow~\cite{AtiyahSingerSkew,CLPS,BCLR}
\begin{equation}\label{eq:realindex}
 \begin{aligned}
 \nu_{\mathrm{cyl}}
 &=\SF_2\{A_{R,t}\}=N_{\mathrm{pair}}\pmod2\\
 &=\frac12\dim_\R\ker L_R\pmod2\\
 &=\frac12\dim_\C\ker\widehat H_\varepsilon\pmod2.
 \end{aligned}
\end{equation}
The real kernel in the middle line is complexified in the last line, giving the module-counting normalization just described. The phase relation is therefore
\begin{equation}\label{eq:sign-theorem}
 \boxed{\mathcal P[Q_1,Q_0]=(-1)^{\nu_{\mathrm{cyl}}}.}
\end{equation}
The positive parameter $\varepsilon$ does not change the protected index. At $r=1,2$ and for the positive-dimensional textures covered by Sec.~\ref{sec:texture-index-isomorphism}, its value is already fixed by the relative homotopy class. Equation~\eqref{eq:sign-theorem} supplies the additional identification with the regulated fermion phase. In the additional paired case $r=5$, each full irreducible kernel module has eight real dimensions, so the half-complex-dimension count is even. This explains why a real-Pfaffian description alone does not imply a protected theta sign. The local crossing model illustrates the theorem but does not replace its Fredholm hypotheses.

\subsection{Applying the index theorem to the Pfaffian sign}
\label{sec:adiabatic-sector-sign}
Let $D>0$ and $r=1,2$, and impose the stable-rank, minimal-module, and slowly varying endpoint hypotheses of Sec.~\ref{sec:texture-index-isomorphism}. Choose a reference texture $Q_*$ and let $\nu_{\mathrm{hom}}[Q]\in\Z_2$ denote the homotopy class of $Q$ relative to $Q_*$, with $\nu_{\mathrm{hom}}[Q_*]=0$. Unlike this reference-based label of one texture, $\nu_{\mathrm{cyl}}[Q_1,Q_0]$ is the quotient index of the cylinder joining two endpoint textures. Since the isomorphism in Eq.~\eqref{eq:physical-index-isomorphism} maps the nonzero class to the nonzero class, it gives
\begin{equation}\label{eq:homotopy-cylinder-parity}
 \boxed{\nu_{\mathrm{cyl}}[Q_1,Q_0]
 =\nu_{\mathrm{hom}}[Q_1]-\nu_{\mathrm{hom}}[Q_0]
 \pmod2.}
\end{equation}
Combining this result with Eq.~\eqref{eq:sign-theorem} identifies the relative Pfaffian sign with relative texture topology. In particular, the sign is unchanged within a homotopy sector and reverses between the two sectors. No further homotopy or additivity assumption is needed here: the uniform gap and the additive physical index map were established in Sec.~\ref{sec:index-well-defined}. The new ingredient of the present section is the phase--index relation, which requires the fixed congruence and the common regularization of Sec.~\ref{sec:congruence-pfaffian}.

\subsection{The resulting \texorpdfstring{$\Z_2$}{Z2} theta weight and microscopic matching}
\label{sec:theta-from-orientation}
With $\nu_{\mathrm{hom}}[Q_*]=0$, Eqs.~\eqref{eq:sign-theorem} and \eqref{eq:homotopy-cylinder-parity} give the nontrivial character directly:
\begin{equation}\label{eq:desired}
 \boxed{\mathcal P_{\mathrm{top}}[Q,Q_*]
 =(-1)^{\nu_{\mathrm{hom}}[Q]}.}
\end{equation}
Writing the general discrete weight as $(-1)^{k\nu_{\mathrm{hom}}[Q]}$, with $k=0,1$, the minimal-model coefficient is therefore fixed:
\begin{equation}\label{eq:theta-generator-index}
 k=\nu_{\mathrm{cyl}}[Q_B,Q_*]=1\pmod2.
\end{equation}
Consequently,
\begin{equation}
 S_{\mathrm{top}}[Q]=i\pi\nu_{\mathrm{hom}}[Q]\pmod{2\pi i}.
\end{equation}
The generator calculation in Sec.~\ref{sec:index-local-calculation} supplies the normalization used in this deduction; it is not a separate calculation still required at this stage. The known two-dimensional AII sign is a check of the result~\cite{RyuAnomaly}. Its Hermitian representative is related to ours by Eq.~\eqref{eq:AII2-spectral}.

The conclusion concerns the specified minimal continuum family and common regulator prescription. Independent physical copies contribute their parities additively, so an even number of identical minimal copies has a trivial combined character. A microscopic regulator may also supply an additional ultraviolet phase or topological counterterm; this must be matched to the microscopic theory rather than absorbed into a reference-orientation convention. These issues do not alter the proven index theorem for the specified family, but they can change the topological weight of a different fermion theory.

At $D=0$ the algebraic congruence and parity normalization remain valid, but the allowed fixed-rank saddle components must be treated directly; the positive-dimensional index theorem is not invoked there. Rapidly varying endpoints, exceptional auxiliary rank, and other self-energy couplings likewise require a separate analysis. Ordinary endpoint eta invariants do not replace the real orientation data and their domain and regulator conditions~\cite{Witten,DaiFreed}.

\section{Wess--Zumino terms in the real classes}
\label{sec:wz-descent}
The integer index theorem fixes the topology of closed textures, whereas a Wess--Zumino (WZ) action is a continuously varying phase in one dimension lower. We derive its universal long-wavelength contribution by combining two inputs: the complex-class anomaly calculation of Ref.~\cite{ZhaoWang2015Response}, applied to the actual complex matrix kernel underlying our Pfaffian, and the normalized real texture index of Sec.~\ref{sec:texture-index-isomorphism}. The former determines the local phase variation; the latter determines its periods and extension ambiguity. We work with based sphere textures, stable auxiliary rank, the specified minimal physical modules, and a common anomaly-compatible regulator. The result is a sphere-normalized WZ level, not a claim about all local terms in the effective action or about arbitrary spacetime topology. We import the complex-parent anomaly result as a separate input; the integer kernel theorem alone would not determine this phase.

\subsection{Dimensions, targets, and the complex parent calculation}
Let $n>0$ be the dimension of the sigma model containing the WZ term and set $D=n+1$. To distinguish the two physical Dirac problems, write
\begin{equation}\label{eq:wz-dimension-shift}
 \begin{gathered}
 r_n=s-n-1,\qquad r_D=r_n-1\pmod8,\\
 r_D=3,7\quad\Longleftrightarrow\quad r_n=4,0.
 \end{gathered}
\end{equation}
At these positions, stable Bott homotopy gives $\pi_n(\mathcal M_s)=0$ and $\pi_D(\mathcal M_s)=\Z$~\cite{Bott1959,RyuReview}. Thus any based field $q:S^n\to\mathcal M_s$ extends to $\widetilde q:B^D\to\mathcal M_s$. All extensions and homotopies are taken at sufficiently large finite auxiliary rank. The $D$-dimensional cylinder index used below does not identify the spectra of the $n$- and $D$-dimensional fermion kernels.

Ignoring the auxiliary reality constraint embeds every non-chiral target into the complex Grassmannian and every chiral target into the unitary group:
\begin{equation}\label{eq:wz-target-inclusion}
 \iota:\mathcal M_s\longrightarrow
 \begin{cases}
 BU,&n\text{ odd, non-chiral},\\
 U,&n\text{ even, chiral}.
 \end{cases}
\end{equation}
Here $BU$ denotes the stable fixed-rank Grassmannian component. The map $\iota$ forgets the antiunitary-induced constraint on the auxiliary field: it retains the same matrix $q$ but regards it as a point in the larger complex target. It is an inclusion of allowed field spaces, not an operator acting on the fermions or a projection of their Hilbert space. In the non-chiral classes it retains Hermiticity and $q^2=1$ while dropping the additional transpose constraint involving $b$. In the chiral classes it retains unitarity and the physical chiral operator $\Gamma$ while dropping that additional constraint, giving a class-AIII parent. Thus neither the gamma matrices nor their complex multiplicities are discarded, and the mass embedding remains exactly that of Sec.~\ref{sec:fermionic-action}.

For non-chiral $n=2j-1$ and chiral $n=2j$, respectively, use the following complex forms, with ordinary matrix trace and exterior products:
\begin{equation}\label{eq:wz-complex-forms}
 \begin{aligned}
 \omega_{2j}^{A}
 &=\frac{\tr_{\mathrm{aux}}\Pi(\dd\Pi)^{2j}}
 {j!(2\pi i)^j},\qquad \Pi=\frac{1+Q}{2},\\
 \omega_{2j+1}^{AIII}
 &=\frac{(-1)^j j!}{(2j+1)!(2\pi i)^{j+1}}
 \tr_{\mathrm{aux}}(Q^{-1}\dd Q)^{2j+1}.
 \end{aligned}
\end{equation}
Their spherical periods are integral and a complex Bott generator has period $\pm1$; orient that generator positively. For the even form this follows from the projector Chern character, and for the odd form from its unitary transgression~\cite{ABS,ChernSimons1974,RyuReview}. In particular, for a Clifford Bott sphere the antisymmetrized volume trace and the sphere volume give unit magnitude in every dimension. The equivalent even-form expression is $\tr Q(\dd Q)^{2j}/[2^{2j+1}j!(2\pi i)^j]$. These formulas fix the trace and factorial normalization explicitly; overall signs are tied to the Dirac orientation. We use the parity-anomaly and chiral-Jacobian mechanisms of Ref.~\cite{ZhaoWang2015Response}, expressing their coefficients in this unit-period convention rather than assuming a trace normalization from the term's name alone.

For one positive complex Dirac block, those calculations give the WZ part of the effective action locally as
\begin{equation}\label{eq:wz-parent-variation}
 \delta S_{\mathrm{WZ}}^{\C}
 =2\pi i\int_{S^n}q^*\bigl(\iota_{\delta q}\omega_D^{\C}\bigr).
\end{equation}
The symbol $\iota_{\delta q}$ denotes contraction with the variation; it is distinct from the target inclusion $\iota$. The underlying parity-anomaly mechanism is explicit in three dimensions in Ref.~\cite{Redlich1984}, while anomalous fermion-measure Jacobians provide the chiral mechanism~\cite{Fujikawa1979}. In the non-chiral calculation, local diagonalization $q=g\Lambda g^{-1}$ produces the projected gauge fields whose parity-odd Chern--Simons terms transgress the projector Chern character. The class-A case is illustrated by disordered Weyl fermions: Zhao and Wang obtain the WZ term through bulk--boundary correspondence~\cite{ZhaoWangWeyl2015}, while Altland and Bagrets derive the parity-odd Chern--Simons action in the three-dimensional disorder field theory~\cite{AltlandBagrets2015}. We use the general-dimensional class-A formulation of Ref.~\cite{ZhaoWang2015Response}. In the chiral calculation, a smooth extension of $q$ removes the chiral mass by field redefinitions; their regulated anomalous Jacobian transgresses the odd Chern character. Opposite Dirac orientations reverse the sign, and independent complex blocks add~\cite{ZhaoWang2015Response}. Neither a global diagonalizing $g$ nor a global logarithm of $q$ is assumed. Local charts give Eq.~\eqref{eq:wz-parent-variation}; its global extension consistency will be proved below.

\subsection{Restriction to the real-class Pfaffian}
The required regulator prescription must preserve the fixed pairing $\B$, reproduce the complex parent's vector-invariant anomaly, and add no independent topological response. Real counterterms renormalizing the magnitude and globally defined nonuniversal local terms are separated from the WZ contribution. This is an explicit prescription for the continuum family, not a proof that every microscopic regulator realizes it. Preserving the vector current fixes the anomaly prescription, as in Bardeen's treatment of vector and axial Ward identities~\cite{Bardeen1969}. In even dimension this choice is important, including the two-dimensional subtlety discussed in Ref.~\cite{ZhaoWang2015Response}.

To connect the two functional integrals, temporarily allow an unconstrained complex-target field $Q$ and define the parent kernel on the \emph{same} complex physical and auxiliary spaces:
\begin{equation}\label{eq:wz-parent-kernel}
 \begin{gathered}
 \D_{\C}[Q]=\gamma^\mu\partial_\mu+\Delta M(Q),\\
 \D[q]=\D_{\C}[\iota(q)],\qquad
 \A[q]=\B^{-1}\D_{\C}[\iota(q)].
 \end{gathered}
\end{equation}
The last kernel is antisymmetric for the allowed real-class $q$. We do not require its extension to arbitrary complex-target $Q$ to remain antisymmetric; the parent calculation uses a determinant, not a Pfaffian on that larger field space.

At a common finite cutoff $\Lambda$, the polynomial identity $\det\A_\Lambda=(\Pf\A_\Lambda)^2$ and the field independence of $\B_\Lambda$ give
\begin{equation}\label{eq:wz-finite-pf-det}
 \begin{gathered}
 (\Pf\A_\Lambda[q])^2
 =\det(\B_\Lambda^{-1})\det\D_{\C,\Lambda}[\iota(q)],\\
 \frac{\delta\Pf\A_\Lambda}{\Pf\A_\Lambda}
 =\frac12\frac{\delta\det\D_{\C,\Lambda}}{\det\D_{\C,\Lambda}}
 =\frac12\Tr(\D_{\C,\Lambda}^{-1}\delta\D_{\C,\Lambda}).
 \end{gathered}
\end{equation}
The second line is evaluated on an invertible patch and along a variation tangent to the real target. In particular, the factor $1/2$ relates \emph{logarithmic} variations, not the unnormalized variations of the Pfaffian and determinant themselves. The constant determinant of $\B_\Lambda^{-1}$ drops out.

Define $S_{\det}^{\C}[Q]=-\log\det_{\mathrm{reg}}\D_{\C}[Q]$ and $S_{\mathrm f}^{\R}[q]=-\log\Pf_{\mathrm{reg}}\A[q]$. After matched regularization, the chain rule yields
\begin{equation}\label{eq:wz-pfaffian-variation}
 \boxed{\begin{aligned}
 \delta S_{\mathrm f}^{\R}[q;\delta q]
 &=\frac12\delta S_{\det}^{\C}
 [\iota(q);\dd\iota_q(\delta q)]\\
 &=-\frac12\Tr_{\mathrm{reg}}
 (\D[q]^{-1}\delta\D[q]).
 \end{aligned}}
\end{equation}
The notation displays the field and its tangent variation; $\dd\iota_q(\delta q)$ is the same matrix variation viewed in the complex target. Equation~\eqref{eq:wz-pfaffian-variation} is the precise sense in which the real-class action variation is half the restricted complex determinant variation. It does not choose a square-root branch independently at each field. A fixed measure and reference phase fix the local integration constant, while the extension ambiguity is checked separately below.

There is one further step before using the minimal complex-class result of Ref.~\cite{ZhaoWang2015Response}. The complex parent retains the full physical representation, which can contain several minimal complex Dirac blocks. Their determinant logarithms add. For the WZ contribution the orientation signs matter: writing $N_\C$ for the signed block count gives
\begin{equation}\label{eq:wz-block-restriction}
 \delta S_{\mathrm{WZ}}^{\R}[q]
 =\frac{N_\C}{2}\,
 \delta S_{\mathrm{WZ},1}^{\C}[\iota(q)],
\end{equation}
where the subscript $1$ denotes one positive complex block and the right-hand variation is along $\dd\iota(\delta q)$. This signed simplification applies to the WZ part, not to all terms of the determinant action. Pulling back the complex anomaly form in Eq.~\eqref{eq:wz-parent-variation} therefore gives
\begin{equation}\label{eq:wz-real-local}
 \delta S_{\mathrm{WZ}}^{\R}
 =2\pi i\,\frac{N_\C}{2}
 \int_{S^n}q^*\bigl(\iota_{\delta q}\iota^*\omega_D^{\C}\bigr).
\end{equation}
The physical block multiplicity has not been absorbed into the trace in Eq.~\eqref{eq:wz-complex-forms}, which acts only on the auxiliary space. This separation avoids counting it twice. The local identity determines the WZ cohomology class; changing a representative by an exact form changes only a globally defined boundary local term.

\subsection{Physical multiplicity and real-target periods}
At $r_n=0,4$, the physical complex Clifford algebra has odd generator number $m=n$ (non-chiral) or $m=n+1$ (chiral), with a single volume sector. Define its signed normalized volume trace by
\begin{equation}\label{eq:wz-physical-block-count}
 N_\C=2^{-(m-1)/2}\tr_{V_{\min}}
 \bigl[(-i)^{(m-1)/2}E_1\cdots E_m\bigr].
\end{equation}
The minimal-module calculation in Sec.~\ref{sec:dirac-minimality} gives $|N_\C|=1$ at $r_n=0$ and $2$ at $r_n=4$. These are complex-block counts, not additional real physical species. Their signs distinguish the two Dirac orientations.

Next evaluate the pulled-back complex form on the real $D$-dimensional texture $Q_B$ constructed in Sec.~\ref{sec:index-texture}. For non-chiral classes it is the Clifford sphere $Q_B=\sum_{a=0}^{D}n_aF_a$; for chiral classes it is $Q_B=n_0+i\sum_{a=1}^{D}n_aF_a$. In both integer cases its auxiliary module occupies one odd volume sector. The antisymmetrized trace in Eq.~\eqref{eq:wz-complex-forms} reduces to the trace of that volume, so it equals $k_B$ times the elementary complex Bott-sphere period. With the positive generator orientation,
\begin{equation}\label{eq:wz-real-period}
 a_{s,D}:=\int_{S^D}Q_B^*\iota^*\omega_D^{\C}
 =k_B=\begin{cases}
 1,&r_D=3\ (r_n=4),\\
 2,&r_D=7\ (r_n=0).
 \end{cases}
\end{equation}
Thus restricting a unit-normalized complex form need not give a unit-normalized real form. The real texture is a generator by the physical index theorem, not by an assumed identification of complex and real charges.

Define a real form, normalized on that spherical generator, by
\begin{equation}\label{eq:wz-normalized-real-form}
 \Omega_D=\frac{1}{a_{s,D}}\iota^*\omega_D^{\C}.
\end{equation}
Orient the integer index positively on $Q_B$. Integration of the closed form and the physical index are both homomorphisms on the sphere-texture group, and both take this generator to one. The theorem therefore gives
\begin{equation}\label{eq:integer-texture}
 \boxed{N[Q]:=\int_{S^D}Q^*\Omega_D
 =\Ind_{\mathfrak C}\widehat H[Q,Q_*]\in\Z.}
\end{equation}
This is the required identification of a differential-form representative with the physical integer index. It is asserted on sphere textures; spherical normalization does not by itself prove integral periods on every cycle of every target.

\begin{table}[tb]\centering\small
\caption{Normalization factors for the real WZ level on a sphere, for one minimal physical module. The Pfaffian contributes $1/2$ in both rows; the Dirac and integer-index orientations fix the sign.}
\label{tab:wz-level-factors}
\begin{tabular}{@{}ccccc@{}}
\toprule
$r_n$&$r_D$&$|N_\C|$&$a_{s,D}$&$|N_\C|a_{s,D}/2$\\\midrule
0&7&1&2&1\\
4&3&2&1&1\\\bottomrule
\end{tabular}
\end{table}

\subsection{The WZ functional and extension independence}
Substitution into Eq.~\eqref{eq:wz-real-local} yields the local variation of
\begin{equation}\label{eq:descent-wz}
 \boxed{\begin{gathered}
 S_{\mathrm{WZ}}[q]=2\pi i\,\ell
 \int_{B^D}\widetilde q^{\,*}\Omega_D,\\
 \ell=\frac{N_\C a_{s,D}}{2}.
 \end{gathered}}
\end{equation}
Indeed, closedness of $\Omega_D$ and Stokes' theorem give
\begin{equation}\label{eq:wz-boundary-variation}
 \delta S_{\mathrm{WZ}}
 =2\pi i\,\ell\int_{S^n}
 q^*\bigl(\iota_{\delta q}\Omega_D\bigr).
\end{equation}
The local anomaly therefore fixes the continuous functional, up to the separated local terms and a reference constant. Table~\ref{tab:wz-level-factors} gives $|\ell|=1$ for one minimal real physical module. Independent physical modules add their signed levels.

Following the extension-quantization construction of the WZ action~\cite{WittenWZ1983}, glue two extension balls with opposite orientations to obtain $Q_{12}:S^D\to\mathcal M_s$. The already established index pairing now gives
\begin{equation}\label{eq:wz-extension-index}
 \boxed{\begin{aligned}
 S_{\mathrm{WZ}}[\widetilde q_1]-S_{\mathrm{WZ}}[\widetilde q_2]
 &=2\pi i\,\ell N[Q_{12}]\\
 &=2\pi i\,\ell\Ind_{\mathfrak C}
 \widehat H[Q_{12},Q_*].
 \end{aligned}}
\end{equation}
Thus $e^{-S_{\mathrm{WZ}}}$ is independent of the extension. In particular, the Pfaffian half in Eq.~\eqref{eq:wz-real-local} produces no half-integer spherical ambiguity: $N_\C a_{s,D}$ is even. On the connected space of based $S^n$ textures, the local variation, its loop periods, and the phase at $q_*$ fix this WZ factor. This completes the sphere-based derivation in the specified anomaly-compatible prescription, rather than merely checking that a WZ term is permitted.

\subsection{Class--dimension pattern, checks, and scope}
Table~\ref{tab:wz-positions} gives the positive-dimensional positions. The physical block count and the real-target period exchange their factors of two between the two integer branches. Under an eight-dimensional shift the normalized Clifford traces are unchanged, so the same derivation repeats with Bott periodicity.

\begin{table}[tb]\centering\small
\caption{Dimensions of real-class WZ models. Entries are residues of the sigma-model dimension $n$ modulo eight; $n>0$, so residue zero first means $n=8$. The texture index used for extension quantization is in dimension $D=n+1$. Every entry has $|\ell|=1$ for one minimal physical module in the spherical normalization of Eq.~\eqref{eq:wz-normalized-real-form}.}
\label{tab:wz-positions}
\begin{tabular}{@{}lccc@{}}
\toprule
Class&Complex parent&$n$ at $r_n=0$&$n$ at $r_n=4$\\\midrule
AI&A&7&3\\
BDI&AIII&0&4\\
D&A&1&5\\
DIII&AIII&2&6\\
AII&A&3&7\\
CII&AIII&4&0\\
C&A&5&1\\
CI&AIII&6&2\\\bottomrule
\end{tabular}
\end{table}

For example, the two-dimensional DIII model has one complex Dirac block, but the vector-trace complex three-form has period two on its real generator. The two-dimensional CI model has two same-orientation complex blocks and period one. In either case the Pfaffian factor gives a unit spherical level. This does not equate the levels quoted in arbitrary group-trace conventions; it specifies precisely the normalization used here.

The adaptation uses the anomaly calculation of the complex parent, not merely the equality of homotopy and index groups. Conversely, the parent calculation alone would leave the real-target periods and the Pfaffian normalization unchecked. Their combination proves Eq.~\eqref{eq:descent-wz} as the universal WZ contribution in the controlled derivative expansion. It does not identify the complete finite-gradient effective action. A different microscopic regulator can add an independent quantized topological response, which must be matched separately; a field-independent reference orientation cannot do so.

Finally, the present proof is for extendible based sphere fields at stable rank. On general manifolds, nonextendible maps, spin-dependent refinements, and torsion or flat phases require a global determinant/Pfaffian-line or anomaly-inflow construction~\cite{DaiFreedOriginal,DaiFreed,YonekuraWZ}. We do not infer those refinements from the integer kernel index or from the differential form alone.

\section{Discussion and conclusions}
\label{sec:status}
We have developed an index-theoretic framework connecting boundary Dirac theories to the topological terms of their disorder nonlinear sigma models. Its two central constructions are the elliptic interpolation cylinder and the quotient index for that cylinder, defined through its graded Clifford symmetry algebra. The index theorem then relates the protected zero-mode index to the relative homotopy class of the endpoint textures. Together, these establish the operator-level correspondence needed to connect texture topology to the underlying fermion integral, rather than merely enumerate the topological terms allowed by the sigma-model target. The subsequent Pfaffian and anomaly calculations turn this correspondence into the $\Z_2$ theta weights and real-class Wess--Zumino terms for the specified minimal models.

The general quotient-index framework for zero-mode stability in the presence of anticommuting symmetries was developed and systematically investigated in the unpublished work of Zheng Zhang and Y.~X.~Zhao~\cite{ZhangZhaoUnpublished}. The present work applies that framework to the cylinder operators, computes their symmetry-dependent quotient indices, and establishes the texture--index correspondence and its connection to the fermionic topological terms. Thus the general zero-mode stability framework is distinct from the cylinder construction and its physical application developed here.

The construction begins with the physical Clifford module of the boundary fermions. Its intertwiners determine the full constant internal graded symmetry algebra of the elliptic interpolation cylinder, retaining the quaternionic multiplicities and volume-sector exchanges required by physical symmetry. The graded algebra depends only on $r=s-D-1\pmod8$, although the physical spinor dimension need not. Its module-extension quotient distinguishes protected zero modes from removable degeneracies: it is $\Z_2$ at $r=1,2$, $\Z$ at $r=3,7$, and trivial otherwise. In particular, the full algebra explains why candidate parities at $r=5,6$ are not protected. This step supplies the representation-theoretic information that target homotopy alone does not determine.

The texture--index theorem identifies the cylinder index with the relative homotopy class of its endpoint textures. Its geometric content is an index-preserving deformation that concentrates the topological obstruction inside a ball. A uniform gap estimate controls sufficiently slowly varying positive-dimensional textures, while gapped collars and real relative-index excision isolate the defect contribution. For the canonical elementary defect, the surrounding sphere carries $Q_B(n)$, and a linear filling gives the solvable operator $H_B$. Its Gaussian zero modes and common Clifford eigenspace determine a unit index at each nonzero entry; additivity then establishes the isomorphism and its normalization. What is preserved throughout is the kernel's index class, not its full dimension. This realizes the correspondence through the cylinder operator rather than inferring it from agreement between abstract groups.

The identification of the fermionic phase is a subsequent step. Section~\ref{sec:congruence-pfaffian} constructs a field-independent congruence from the original complex antisymmetric kernel to a Hermitian antisymmetric representative. Multiplication by $-i$ gives a real skew-adjoint operator, whose Pfaffian has the same relative phase under the common regularization prescription. The congruence supplies the real structure; the regulator and reference orientation define the relative Pfaffian. Section~\ref{sec:paired-spectral-flow} then uses Witten's spectral-flow argument to identify Pfaffian sign transport with the mod-two index. Combining this relation with the texture--index theorem gives, for $D>0$ and $r=1,2$ under the stated Fredholm and regulator assumptions,
\begin{equation}\label{eq:summary-phase-index}
 \boxed{\begin{aligned}
 \mathcal P[Q_1,Q_0]
 &=(-1)^{\nu_{\mathrm{cyl}}[Q_1,Q_0]}\\
 &=(-1)^{\nu_{\mathrm{hom}}[Q_1]-\nu_{\mathrm{hom}}[Q_0]}.
 \end{aligned}}
\end{equation}
Thus one minimal physical module generates the nontrivial $\Z_2$ theta character, not merely a theory in which such a term is allowed. The $r=5$ examples show why the congruence alone is insufficient: a real-Pfaffian description can exist even when the full symmetry makes its mod-two index trivial. The sign transport uses Witten's orientation-flow mechanism without requiring distinct sigma textures to be gauge-equivalent.

The integer indices have a complementary role in the Wess--Zumino terms one dimension lower. As derived in Sec.~\ref{sec:wz-descent}, the complex-parent anomaly determines the local variation, whereas the integer texture index fixes the periods and extension ambiguity. The Pfaffian factor of one half, the signed physical complex-block multiplicity, and the real-target period combine to give a unit-magnitude spherical level for one minimal physical module. Matching the normalized form to the integer index makes the exponentiated action independent of the extension. The anomaly and index calculations are therefore complementary ingredients, rather than alternative ways of inferring the same coefficient from target topology.

These results concern the specified continuum Dirac--$Q$ families at stable auxiliary rank, with sufficiently slowly varying positive-dimensional textures and compatible operator domains, Fredholm realizations, and regularization prescriptions. Applying them to a microscopic system requires the physical species count and any additional ultraviolet topological response; such a response cannot be absorbed into a reference-orientation convention. Other self-energy couplings or exceptional auxiliary ranks require a renewed symmetry analysis, and algebraic zero-mode lifting need not be realizable by a local $Q$-deformation. Fixed-rank components at $D=0$ must be treated separately. The Wess--Zumino derivation is sphere-normalized and does not determine arbitrary finite-gradient terms or the spin, torsion, and nonextendible-field refinements needed on general manifolds.

Within this scope, the framework makes the connection between physical Clifford representations, boundary texture topology, and fermionic topological phases explicit. It provides a common operator-level route from boundary Dirac theories to the sigma-model topological terms associated with the periodic table, connecting the momentum-space and localization perspectives through the underlying fermion integral.

\begin{acknowledgments}
The construction of the cylinder operator and its quotient index, the proof of the texture--index theorem, and the application of Witten's argument to derive the $\Z_2$ theta terms are due to the author. The author used GPT 6 Astra through interactive discussions to assist in writing the preparatory material in Secs.~\ref{sec:minimal-boundary} and \ref{sec:fermionic-action}. GPT 6 Astra also assisted with language polishing throughout the manuscript.

This work was supported by the General Research Fund of Hong Kong (Grant Nos.~17302525 and 17301224).
\end{acknowledgments}

\patchcmd{\bibsection}{\begingroup}{\vspace{1em}\begingroup}{}{}
\bibliography{sigma_model_index_prb}

@article{SRFL,
 author = {Schnyder, Andreas P. and Ryu, Shinsei and Furusaki, Akira and Ludwig, Andreas W. W.},
 title = {Classification of topological insulators and superconductors in three spatial dimensions},
 journal = {Phys. Rev. B}, volume = {78}, pages = {195125}, year = {2008},
 doi = {10.1103/PhysRevB.78.195125}
}

@article{RyuReview,
 author = {Ryu, Shinsei and Schnyder, Andreas P. and Furusaki, Akira and Ludwig, Andreas W. W.},
 title = {Topological insulators and superconductors: tenfold way and dimensional hierarchy},
 journal = {New J. Phys.}, volume = {12}, pages = {065010}, year = {2010},
 doi = {10.1088/1367-2630/12/6/065010}
}

@article{Kitaev,
 author = {Kitaev, Alexei},
 title = {Periodic table for topological insulators and superconductors},
 journal = {AIP Conf. Proc.}, volume = {1134}, pages = {22--30}, year = {2009},
 doi = {10.1063/1.3149495}
}

@article{RyuAnomaly,
 author = {Ryu, Shinsei and Mudry, Christopher and Obuse, Hideaki and Furusaki, Akira},
 title = {{$\mathbb Z_2$} topological term, the global anomaly, and the two-dimensional symplectic symmetry class of {Anderson} localization},
 journal = {Phys. Rev. Lett.}, volume = {99}, pages = {116601}, year = {2007},
 doi = {10.1103/PhysRevLett.99.116601}
}

@article{Koenig,
 author = {K{\"o}nig, E. J. and Ostrovsky, P. M. and Protopopov, I. V. and Mirlin, A. D.},
 title = {Metal-insulator transition in two-dimensional random fermion systems of chiral symmetry classes},
 journal = {Phys. Rev. B}, volume = {85}, pages = {195130}, year = {2012},
 doi = {10.1103/PhysRevB.85.195130}
}

@article{Witten,
 author = {Witten, Edward}, title = {Fermion path integrals and topological phases},
 journal = {Rev. Mod. Phys.}, volume = {88}, pages = {035001}, year = {2016},
 doi = {10.1103/RevModPhys.88.035001}
}

@article{DaiFreed,
 author = {Yonekura, Kazuya}, title = {{Dai--Freed} theorem and topological phases of matter},
 journal = {J. High Energy Phys.}, volume = {2016}, number = {9}, pages = {022}, year = {2016},
 doi = {10.1007/JHEP09(2016)022}
}

@article{YonekuraWZ,
 author = {Yonekura, Kazuya}, title = {General anomaly matching by {Goldstone} bosons},
 journal = {J. High Energy Phys.}, volume = {2021}, number = {3}, pages = {057}, year = {2021},
 doi = {10.1007/JHEP03(2021)057}
}

@article{CLPS,
 author = {Carey, Alan L. and Phillips, John and Schulz-Baldes, Hermann},
 title = {Spectral flow for skew-adjoint {Fredholm} operators},
 journal = {J. Spectr. Theory}, volume = {9}, number = {1}, pages = {137--170}, year = {2019},
 doi = {10.4171/JST/243}, eprint = {1604.06994}, archivePrefix = {arXiv}, primaryClass = {math.FA}
}

@article{BCLR,
 author = {Bourne, Chris and Carey, Alan L. and Lesch, Matthias and Rennie, Adam},
 title = {The {KO}-valued spectral flow for skew-adjoint {Fredholm} operators},
 journal = {J. Topol. Anal.}, volume = {14}, number = {2}, pages = {505--556}, year = {2022},
 doi = {10.1142/S1793525320500557}, eprint = {1907.04981}, archivePrefix = {arXiv}, primaryClass = {math.KT}
}

@article{ABS,
 author = {Atiyah, M. F. and Bott, R. and Shapiro, A.}, title = {Clifford modules},
 journal = {Topology}, volume = {3}, number = {Supplement 1}, pages = {3--38}, year = {1964},
 doi = {10.1016/0040-9383(64)90003-5}
}

@article{WittenSU2,
 author = {Witten, Edward},
 title = {An {SU(2)} anomaly},
 journal = {Phys. Lett. B}, volume = {117}, pages = {324--328}, year = {1982},
 doi = {10.1016/0370-2693(82)90728-6}
}

@article{ZhaoWang2014,
 author = {Zhao, Y. X. and Wang, Z. D.},
 title = {Topological connection between the stability of {Fermi} surfaces and topological insulators and superconductors},
 journal = {Phys. Rev. B}, volume = {89}, pages = {075111}, year = {2014},
 doi = {10.1103/PhysRevB.89.075111}
}

@article{BunkeRelative,
 author = {Bunke, Ulrich},
 title = {A {K}-theoretic relative index theorem and {Callias}-type {Dirac} operators},
 journal = {Math. Ann.}, volume = {303}, pages = {241--279}, year = {1995},
 doi = {10.1007/BF01460989}
}

@article{ZhaoWang2015Response,
 author = {Zhao, Y. X. and Wang, Z. D.},
 title = {General response theory of topologically stable {Fermi} points and its implications for disordered cases},
 journal = {Phys. Rev. B}, volume = {92}, pages = {085143}, year = {2015},
 doi = {10.1103/PhysRevB.92.085143}
}

@article{Anderson1958,
 author = {Anderson, P. W.}, title = {Absence of diffusion in certain random lattices},
 journal = {Phys. Rev.}, volume = {109}, pages = {1492--1505}, year = {1958},
 doi = {10.1103/PhysRev.109.1492}
}

@article{KaneMeleQSH,
 author = {Kane, C. L. and Mele, E. J.}, title = {Quantum spin {Hall} effect in graphene},
 journal = {Phys. Rev. Lett.}, volume = {95}, pages = {226801}, year = {2005},
 doi = {10.1103/PhysRevLett.95.226801}
}

@article{KaneMeleZ2,
 author = {Kane, C. L. and Mele, E. J.}, title = {{$\mathbb Z_2$} topological order and the quantum spin {Hall} effect},
 journal = {Phys. Rev. Lett.}, volume = {95}, pages = {146802}, year = {2005},
 doi = {10.1103/PhysRevLett.95.146802}
}

@article{AltlandZirnbauer,
 author = {Altland, Alexander and Zirnbauer, Martin R.},
 title = {Nonstandard symmetry classes in mesoscopic normal-superconducting hybrid structures},
 journal = {Phys. Rev. B}, volume = {55}, pages = {1142--1161}, year = {1997},
 doi = {10.1103/PhysRevB.55.1142}
}

@article{Zirnbauer1996,
 author = {Zirnbauer, Martin R.},
 title = {Riemannian symmetric superspaces and their origin in random-matrix theory},
 journal = {J. Math. Phys.}, volume = {37}, pages = {4986--5018}, year = {1996},
 doi = {10.1063/1.531675}
}

@article{Wegner1979,
 author = {Wegner, Franz}, title = {The mobility edge problem: Continuous symmetry and a conjecture},
 journal = {Z. Phys. B}, volume = {35}, pages = {207--210}, year = {1979},
 doi = {10.1007/BF01319839}
}

@article{Bott1959,
 author = {Bott, Raoul}, title = {The stable homotopy of the classical groups},
 journal = {Ann. Math.}, volume = {70}, number = {2}, pages = {313--337}, year = {1959},
 doi = {10.2307/1970106}
}

@article{AtiyahReality,
 author = {Atiyah, M. F.}, title = {{K}-theory and reality},
 journal = {Q. J. Math.}, volume = {17}, number = {1}, pages = {367--386}, year = {1966},
 doi = {10.1093/qmath/17.1.367}
}

@article{FuKaneMele3D,
 author = {Fu, Liang and Kane, C. L. and Mele, E. J.},
 title = {Topological insulators in three dimensions},
 journal = {Phys. Rev. Lett.}, volume = {98}, pages = {106803}, year = {2007},
 doi = {10.1103/PhysRevLett.98.106803}
}

@article{Ostrovsky2007,
 author = {Ostrovsky, P. M. and Gornyi, I. V. and Mirlin, A. D.},
 title = {Quantum criticality and minimal conductivity in graphene with long-range disorder},
 journal = {Phys. Rev. Lett.}, volume = {98}, pages = {256801}, year = {2007},
 doi = {10.1103/PhysRevLett.98.256801}
}

@article{AtiyahSingerSkew,
 author = {Atiyah, M. F. and Singer, I. M.},
 title = {Index theory for skew-adjoint {Fredholm} operators},
 journal = {Publ. Math. Inst. Hautes {\'E}tudes Sci.}, volume = {37}, pages = {5--26}, year = {1969},
 doi = {10.1007/BF02684885}
}

@article{Takagi1924,
 author = {Takagi, Teiji},
 title = {On an algebraic problem related to an analytic theorem of {Carath\'eodory} and {Fej\'er} and on an allied theorem of {Landau}},
 journal = {Jpn. J. Math.}, volume = {1}, pages = {83--93}, year = {1924},
 doi = {10.4099/jjm1924.1.0_83}
}

@article{WessZumino1971,
 author = {Wess, J. and Zumino, B.}, title = {Consequences of anomalous {Ward} identities},
 journal = {Phys. Lett. B}, volume = {37}, pages = {95--97}, year = {1971},
 doi = {10.1016/0370-2693(71)90582-X}
}

@article{WittenWZ1983,
 author = {Witten, Edward}, title = {Global aspects of current algebra},
 journal = {Nucl. Phys. B}, volume = {223}, pages = {422--432}, year = {1983},
 doi = {10.1016/0550-3213(83)90063-9}
}

@article{Fujikawa1979,
 author = {Fujikawa, Kazuo}, title = {Path-integral measure for gauge-invariant fermion theories},
 journal = {Phys. Rev. Lett.}, volume = {42}, pages = {1195--1198}, year = {1979},
 doi = {10.1103/PhysRevLett.42.1195}
}

@article{Redlich1984,
 author = {Redlich, A. N.}, title = {Parity violation and gauge noninvariance of the effective gauge field action in three dimensions},
 journal = {Phys. Rev. D}, volume = {29}, pages = {2366--2374}, year = {1984},
 doi = {10.1103/PhysRevD.29.2366}
}

@article{Bardeen1969,
 author = {Bardeen, William A.}, title = {Anomalous {Ward} identities in spinor field theories},
 journal = {Phys. Rev.}, volume = {184}, pages = {1848--1859}, year = {1969},
 doi = {10.1103/PhysRev.184.1848}
}

@article{ChernSimons1974,
 author = {Chern, Shiing-Shen and Simons, James}, title = {Characteristic forms and geometric invariants},
 journal = {Ann. Math.}, volume = {99}, number = {1}, pages = {48--69}, year = {1974},
 doi = {10.2307/1971013}
}

@article{DaiFreedOriginal,
 author = {Dai, Xianzhe and Freed, Daniel S.}, title = {Eta invariants and determinant lines},
 journal = {J. Math. Phys.}, volume = {35}, pages = {5155--5194}, year = {1994},
 doi = {10.1063/1.530747}, eprint = {hep-th/9405012}, archivePrefix = {arXiv}
}

@article{AtiyahSingerReal,
 author = {Atiyah, M. F. and Singer, I. M.},
 title = {The index of elliptic operators: {V}},
 journal = {Ann. Math.}, volume = {93}, number = {1}, pages = {139--149}, year = {1971},
 doi = {10.2307/1970757}
}

@article{Pruisken1984,
 author = {Pruisken, A. M. M.},
 title = {On localization in the theory of the quantized {Hall} effect: A two-dimensional realization of the {$\theta$}-vacuum},
 journal = {Nucl. Phys. B}, volume = {235}, pages = {277--298}, year = {1984},
 doi = {10.1016/0550-3213(84)90101-9}
}

@article{PruiskenEdge1999,
 author = {Pruisken, A. M. M. and {\v S}kori{\'c}, B. and Baranov, M. A.},
 title = {(Mis-)handling gauge invariance in the theory of the quantum {Hall} effect. {III}. The instanton vacuum and chiral-edge physics},
 journal = {Phys. Rev. B}, volume = {60}, pages = {16838--16864}, year = {1999},
 doi = {10.1103/PhysRevB.60.16838}
}

@article{Pruisken2009,
 author = {Pruisken, A. M. M.},
 title = {Super universality of the quantum {Hall} effect and the ``large {$N$} picture'' of the {$\vartheta$} angle},
 journal = {Int. J. Theor. Phys.}, volume = {48}, pages = {1736--1765}, year = {2009},
 doi = {10.1007/s10773-009-9947-7}
}

@article{ZhaoWang2013,
 author = {Zhao, Y. X. and Wang, Z. D.},
 title = {Topological classification and stability of {Fermi} surfaces},
 journal = {Phys. Rev. Lett.}, volume = {110}, pages = {240404}, year = {2013},
 doi = {10.1103/PhysRevLett.110.240404}
}

@article{MorimotoFurusaki2013,
 author = {Morimoto, Takahiro and Furusaki, Akira},
 title = {Topological classification with additional symmetries from {Clifford} algebras},
 journal = {Phys. Rev. B}, volume = {88}, pages = {125129}, year = {2013},
 doi = {10.1103/PhysRevB.88.125129}
}

@article{ZhaoWangWeyl2015,
 author = {Zhao, Y. X. and Wang, Z. D.},
 title = {Disordered {Weyl} semimetals and their topological family},
 journal = {Phys. Rev. Lett.}, volume = {114}, pages = {206602}, year = {2015},
 doi = {10.1103/PhysRevLett.114.206602}
}

@article{AltlandBagrets2015,
 author = {Altland, Alexander and Bagrets, Dmitry},
 title = {Effective field theory of the disordered {Weyl} semimetal},
 journal = {Phys. Rev. Lett.}, volume = {114}, pages = {257201}, year = {2015},
 doi = {10.1103/PhysRevLett.114.257201}
}

@misc{ZhangZhaoUnpublished,
 author = {Zhang, Zheng and Zhao, Y. X.},
 note = {Unpublished work}
}
\end{document}